\documentclass[twocolumn]{autart}
\pdfoutput=1
\usepackage[T1]{fontenc}
\usepackage{amsmath}
\usepackage{psfrag}
\usepackage{amssymb}
\usepackage{color}
\usepackage{multirow}
\usepackage{url}
\DeclareMathOperator*{\argmin}{argmin}

\newcommand{\diag}{{\mathrm{diag}}}
\newcommand{\adj}{{\mathrm{adj}}}
\renewcommand{\Re}{{\mathbb{R}}}
\newcommand{\Co}{{\mathbb{C}}}
\newcommand{\BEQ}{\begin{equation}}
\newcommand{\EEQ}{\end{equation}}
\newcommand{\reals}{{\mathbb{R}}}
\newcommand{\symm}{{\mathbb{S}}}
\newcommand{\Aa}{\mathcal A_\mathrm{adj}}
\newcommand{\BEA}{\begin{eqnarray}}
\newcommand{\EEA}{\end{eqnarray}}
\newcommand{\QED}{\hfill \ensuremath{\Box}}

\newtheorem{theorem}{Theorem}
\newtheorem{definition}[thm]{Definition}
\newtheorem{corollary}[thm]{Corollary}
\newtheorem{lemma}[thm]{Lemma}
\DeclareMathOperator*{\Tr}{Tr}
\newtheorem{remark}[thm]{Remark}

\usepackage{graphicx}      
\usepackage{natbib}        
\begin{document}
\begin{frontmatter}

\title{N2SID: Nuclear Norm Subspace Identification \thanksref{footnoteinfo}} 

\thanks[footnoteinfo]{Part of the research was done while the first author
 was a Visiting Professor at the
Division of Automatic Control, Department of Electrical Engineering, 
Link\"oping University, Sweden.
This work was partially supported by the European Research
Council Advanced Grant Agreement No. 339681. Corresponding Author: 
{\tt\small  m.verhaegen@tudelft.nl.}}

\author[First]{Michel Verhaegen} 
\author[Second]{Anders Hansson} 

\address[First]{Delft Center for Systems and
  Control\\ Delft University\\Delft, The Netherlands}
\address[Second]{Division of Automatic Control\\
 Link\"oping University\\ Link\"oping, Sweden}

\begin{abstract}                
The identification of multivariable state space models in 
innovation form is solved in a subspace identification framework using
convex nuclear norm optimization. The convex optimization approach
allows to include constraints on the unknown matrices in the
data-equation characterizing subspace identification methods, such as
the lower triangular block-Toeplitz of weighting matrices constructed
from the Markov parameters of the unknown observer.   The
classical use of instrumental variables to remove the influence of the
innovation term on the data equation in subspace identification is
avoided. The avoidance of the instrumental
variable projection step has the potential to improve the accuracy of
the estimated model predictions, especially for short data length sequences. This is illustrated using a data set
from the DaSIy library. An efficient implementation in the framework
of the Alternating Direction Method of Multipliers (ADMM) is presented
that is used in the validation study. 
\end{abstract}


\end{frontmatter}

\section{Introduction}

The generation of Subspace IDentification (SID) 
methods for the identification of Linear
Time-Invariant (LTI) state space models as developed originally in
\cite{OvS94,Lar90,Ver94}  derive \emph{approximate}
models rather than models that are ``optimal'' with respect to a
goodness of fit criterium defined in terms of the weighted norm of the
difference
between the measured output and the model predicted output. 
The approximation is based on linear algebra
transformations and factorizations of structured Hankel matrices
constructed from the input-output data that are related via the
so-called {\em data equation} \cite{VerBook}. All existing SID methods aim to derive a low rank matrix
from which key subspaces, hence the name subspace identification, are
derived. The low rank approximation is in general done using a
Singular Value Decomposition (SVD). 

A number of recent developments have been made to integrate the low
rank approximation step in SID with a goodness of fit into a single
multi-criteria convex optimization problem. These contributions were
inspired by the work in \cite{Faz:02}  to approximate a
constraint on the rank of a matrix by
minimizing its nuclear norm. It resulted  into a number of improvements to
the low rank approximation step over the classically
used SVD  in SID,
\cite{Liu+van09,LiV:09,MoF:10,faz+pon+sun+tse12,HLV:12,liu+han+van13,Smith14,Smith12}. 

When considering identifying innovation state space models, a common
approach is to make use of instrumental variables 
\cite{HLV:12} . It is well known
that the projection operation related to the use of instrumental
variables may result into a degradation of the accuracy of the estimated
quantities. 

In this paper we present a new SID method  for identifying
multivariable state space models in 
innovation form
within the
framework of nuclear norm optimization. The new SID method avoids the use of
instrumental variables.  The method is a convex relaxation of
Pareto optimization in which structural constraints are
imposed on the unknowns in the data
equation, such as their block-Toeplitz matrix structure. This Pareto
optimization approach allows to
make a trade-off between a Prediction Error type of optimality
criteria, that is minimizing the (co-)variance of the one-step ahead
prediction of a linear Kalman filter type observer, on one hand, and finding an observer
of lowest complexity, i.e. of lowest model order, on the other hand. A
key result is that the
structural Toeplitz constraint is sufficient to find the minimal
observer realization when the optimal one-step ahead prediction of the
output is known. The incentive to estimate a Kalman filter type of
observer  also justifies the constraint to attempt to minimize the
variance of the one-step ahead prediction error. \\[3pt]
It is interesting to note that this key result stipulates precise
conditions on the persistancy of excitation of the input (in open-loop
experiments). For many instrumental variable based SID methods it is
still an open question what the persistency of excitation condition is on
a generic input sequence to guarantee the algorithm to work for finite
data length samples or to be consistent \cite{Jansson98}.   

The convex relaxation of the new SID
approach is denoted by Nuclear Norm Subspace IDentification (N2SID). 
Its advantages are demonstrated in a comparison study on real-life data sets in the 
 DaSIy data base, \cite{DDDF:97}. In this comparison study N2SID
 demonstrated to yield models that lead to improved predictions 
over two existing 
SID methods, N4SID and the recent Nuclear Norm based SID methods
presented in \cite{liu+han+van13} and with the Prediction Error Method
(PEM) \cite{Ljungtb:07}. For the sake of
brevity the results about only one data set are reported. For more
information on a more extensive experimental analysis leading to
similar conclusions we refer to \cite{Report}. 

The foundations for N2SID were presented in 
\cite{DBLP:journals/corr/VerhaegenH14}. There
the resulting optimization problem was solved using a Semi-Definite 
Programming (SDP) solver after a reformulation of the problem into an
equivalent SDP problem. In addition to this the problem formulation was
approximated in order to obtain a problem of manageable size for current
SDP solvers. In this paper we will instead solve the problem with 
the Alternating Direction Method of Multipliers (ADMM). ADMM is known to 
be a good choice for solving regularized nuclear norm problems as the ones
we are solving in this paper, \cite{liu+han+van13}. In order to get an
efficient implementation we have customized the computations for obtaining
the coefficient matrix of the normal equations associated with our formulation
using Fast Fourier Transformation (FFT) techniques. This is the key to obtain
an efficient implementation. 

The paper is organized as follows. In Section~2 the identification
problem for identifying a multi-variable state space model in  a
subspace context while taking a prediction error cost function into
consideration is presented. The data equation and necessary
preliminaries on the assumptions made in the
analysis and description of the subspace identification method is
presented in Section~3. The multi-criteria optimization problem, the
analysis of the uniqueness of solution and
its convex relaxation are presented in Section~4. In Section~5 we explain how to obtain an efficient ADMM code.
The results of the performances are
illustrated in Section~6 in a comparison study of N2SID with two other
SID methods, N4SID and the recent Nuclear Norm based SID methods
presented in \cite{liu+han+van13} and with PEM. 
Finally, we end this paper with some
concluding remarks. 
\subsection{Notations}
We introduce the Matlab-like notation that for a vector or matrix 
$X\in\Re^{M\times N}\left(\Co^{M\times N}\right)$ it holds
that $X_{m:n,p:q}$ is the sub-matrix of $X$ with rows $m$ through $n$ and columns
$p$ through $q$. If one of the dimensions of the matrix is $n$, then 
$1:n$ can be abbreviated as just $:$. Moreover, with $X_{m:-1:n,p:q}$ is
meant the sub-matrix of $X$ obtained by taking rows $m$ through $n$ in 
reverse order, where $m$ is greater than or equal to $n$. Similar notation
may be used for the columns.  
\section{The Subspace Identification Problem}
\label{sec:S2IDE}

In system identification a challenging problem is to identify Linear
Time Invariant (LTI) systems with multiple inputs and multiple outputs
using short length data sequences. Taking process and measurement
noise into consideration, a general state space model for LTI systems can be given in
so-called innovation form, \cite{VerBook}:
\begin{equation}
\label{ssmI}
\left\{  \begin{array}{rcl}
x(k+1) & = & A x(k) + B u(k) + K e(k) \\
y(k) & = & C x(k) + D u(k) + e(k)
\end{array} \right.
 \end{equation}
with $x(k) \in \mathbb{R}^n, y(k) \in \mathbb{R}^p, u(k) \in
\mathbb{R}^m$ and $e(k)$ a zero-mean white noise sequence. 
Since we are interested in short data sets no requirement on
consistency is included in the following problem formulation. 

{\em Problem Formulation:} 
Given the input-ouput (i/o) data batches $\{u(k),y(k)\}_{k=1}^N$, with
$N > n$ and assumed to be retrieved from an identification experiment
with a system belonging to the class of LTI systems as
represented by \eqref{ssmI}, the problem is 
to determine approximate system
matrices $(\hat{A}_T,\hat{B}_T,\hat{C}_T,\hat{D},\hat{K}_T)$ that
define  the $\hat{n}$-th order  observer of {\em ``low''} complexity:
 \begin{equation}
\label{ssmOT}
\left\{  \begin{array}{rcl}
 \hat{x}_T(k+1) & = & \hat{A}_T \hat{x}_T(k) + \hat{B}_T u_v(k) + \hat{K}_T \Big( y_v(k) - \hat{C}_T \hat{x}_T \Big)  \\
\hat{y}_v(k) & = & \hat{C}_T \hat{x}_T(k) + \hat{D} u_v(k) 
\end{array} \right.
 \end{equation}
such that the approximated output $\hat{y}_v(k)$ is {\em ``close''} to the
measured output $y_v(k)$ of the validation pair $\{ u_v(k),y_v(k)
\}_{k=1}^{N_v}$ as expressed by a small value of the cost function,
\begin{equation}
 \label{CostF}
 \frac{1}{N_v} \sum_{k=1}^{N_v} \| y_v(k) - \hat{y}_v(k) \|_2^2.
\end{equation}
The quantitative notions like {\em ``low''} and {\em ``close approximation''} will be
made more precise in the new N2SID solution toward this problem. The
solution to this problem is provided under 2 Assumptions. The first is
listed here, the second at the end of Section~\ref{s_dataeq}.

{\bf Assumption A.1.} The pair $(A,C)$ is observable and the pair $(A,
\left[ \begin{matrix} B & K \end{matrix} \right])$ is reachable. 

A key starting point in the formulation of subspace methods is the
relation between structured Hankel matrices constructed from the i/o
data. This relationship  will as defined in \cite{VerBook} be
the data equation. It will be presented in the next
section. 

\section{The Data Equation, its structure and Preliminaries}
 \label{s_dataeq}
Let the LTI model \eqref{ssmI} be represented in its so-called
observer form:
\begin{equation}
\label{ssmO}
\left\{  \begin{array}{rcl}
x(k+1) & = & (A - KC) x(k) + (B - KD) u(k) + K y(k) \\
y(k) & = & C x(k) + D u(k) + e(k)
\end{array} \right.
 \end{equation}
We will denote this model compactly as:
\begin{equation}
\label{ssmOc}
\left\{  \begin{array}{rcl}
x(k+1) & = & \mathcal{A} x(k) + \mathcal{B} u(k) + K y(k) \\
y(k) & = & C x(k) + D u(k) + e(k)
\end{array} \right.
 \end{equation}
with $\mathcal{A}$ the observer system matrix $(A-KC)$ and
$\mathcal{B}$ equal to $(B-KD)$. Though this property will not be used
in the sequel, the matrix $\mathcal{A}$ can be assumed to be
asymptotically stable.

For the construction of the data equation, we store the measured i/o
data in block-Hankel matrices. For fixed $N$ assumed to be larger
then the order $n$ of the underlying system, the definition of the
number of block-rows fully defines the size of these Hankel
matrices. Let this dimensioning parameter be denoted by $s$, then the
Hankel matrix of the input is defined as:
\begin{equation}
 \label{defHaki}
U_{s,N} = \left[ \begin{matrix} u(1) & u(2) & \cdots & u(N-s+1) \\ u(2) &
    u(3) &   & \vdots \\ \vdots &  & \ddots &  \\ u(s) & u(s+1) &
    \cdots &  u(N) \end{matrix} \right].
\end{equation}
The Hankel matrices from the output $y(k)$ and the innovation
$e(k)$ are defined similarly and
denoted by $Y_{s,N}$ and $E_{s,N}$, respectively. The relationship between
these Hankel matrices, that readily follows from the linear model
equations in \eqref{ssmOc}, require the definition of the following
\emph{structured} matrices. First we define the extended observability matrix
$\mathcal{O}_s$:
\begin{equation}
 \label{Osv}
\mathcal{O}_s^T = \left[ \begin{matrix} C^T & \mathcal{A}^T C^T & \cdots &
    \mathcal{A^T}^{s-1} C^T \end{matrix} \right].
\end{equation}
Second, we define a Toeplitz matrix from the quadruple of systems
matrices $\{\mathcal{A},\mathcal{B},C,D\}$ as:
\begin{equation}
 \label{Toepl}
T_{u,s} = \left[ \begin{matrix} D & 0 & \cdots & 0 \\ C
    \mathcal{B} & D &   & 0 \\ \vdots &  & \ddots & \\
C \mathcal{A}^{s-2} \mathcal{B} &  & \cdots & D  \end{matrix} \right] 
\end{equation}
and in the same way we define a Toeplitz matrix $T_{y,s}$ from the
quadruple $\{\mathcal{A},K,C,0\}$. Finally, let the state sequence be
  stored as:
\begin{equation}
 \label{state}
X_N = \left[ \begin{matrix} x(1) & x(2) & \cdots & x(N-s+1) \end{matrix}
\right].
\end{equation}
Then the data equation compactly reads:
\begin{equation}
 \label{DataEq}
Y_{s,N} = \mathcal{O}_s X_N + T_{u,s} U_{s,N} + T_{y,s} Y_{s,N} + E_{s,N}.
\end{equation}
This equation is a simple linear matrix equation that highlights the
challenges in subspace identification, which is to approximate from the
given Hankel matrices $Y_{s,N}$ and $U_{s,N}$ the column space of the
observability matrix and/or that of the state sequence of the observer~\eqref{ssmOc}.

The equation is highly structured. In this
paper we focus on the following key structural properties about the
unknown matrices in \eqref{DataEq}:
\begin{enumerate}
 \item The matrix product $\mathcal{O}_s X_N$ is \emph{low rank} when $s >
   n$. 
 \item The matrices $T_{u,s}$ and $T_{y,s}$ are block-Toeplitz.
 \item The matrix $E_{s,N}$ is block-Hankel.
\end{enumerate}
The interesting observation is that these 3 structural properties can
be added as constraints to the multi-criteria optimization problem
considered while preserving convexity. This is demonstrated in Section
\ref{sec4}. 

The analysis in Section \ref{sec4}  requires the following
preliminaries. 

\begin{definition} \cite{VerBook}: 
 \label{def:pe}
 A signal $u(k) \in \mathbb{R}^m$ is
  persistently exciting of order $s$ if and only if there exists an
  integer $N$ such that the matrix $U_{s,N}$ has full row rank. 
\end{definition}

\begin{lemma} \cite{Jansson98}: Consider the state space model in
  innovation form \eqref{ssmI} and let all stochastic signals be
  stationary and ergodic, let Assumption A.1 be satisfied and let the input $u(k)$ be quasi-stationary \cite{Ljungtb:07} and 
  persistently exciting of order $s+n$, then:
\[
 \lim_{N \rightarrow \infty} \frac{1}{N} \left[ \begin{matrix} X_N \\
     U_{s,N} \end{matrix} \right] \left[ \begin{matrix} X_N^T &
     U_{s,N}^T \end{matrix} \right] > 0
\]
 \label{lem:Jansson} 
\end{lemma}

\begin{corollary} \label{cor1}
Let the conditions stipulated in Lemma~\ref{lem:Jansson} hold,  and
let $u(k)$ be statistically independent from the innovation sequence
$e(\ell)$ for all $k, \ell$, then,
\[
 \lim_{N \rightarrow \infty} \frac{1}{N} \left[ \begin{matrix} X_N \\
     U_{s,N} \\ Y_{s,N} \end{matrix} \right] \left[ \begin{matrix} X_N^T &
     U_{s,N}^T & Y_{s,N}^T \end{matrix} \right] > 0
\]
\end{corollary}

{\bf Proof:}  Since $e(k)$ is white noise, it follows that $\mathbb{E}[ x(k)
e(\ell)^T] = 0$ (with $\mathbb{E}$ denoting the expectation operator), for
$\ell \geq k$. This in combination with the independency between
$u(k)$ and $e(\ell)$, the white noise property of $e(k)$ and the
ergodicity or the quasi-stationarity of the signals yields,
\begin{equation}
 \label{posdef}
 \lim_{N \rightarrow \infty} \frac{1}{N} \left[ \begin{matrix} X_N \\
     U_{s,N} \\ E_{s,N} \end{matrix} \right] \left[ \begin{matrix} X_N^T &
     U_{s,N}^T & E_{s,N}^T \end{matrix} \right] > 0
\end{equation}
Considering model~\eqref{ssmI}, let the block-Toeplitz matrices $T_{u,s}^\prime$
and $T_{e,s}$ be defined as the Toeplitz matrix $T_{u,s}$ in
\eqref{Toepl} but from the quadruples $(A,B,C,D)$ and $(A,K,C,I)$,
respectively. Let $O_s^T =
\left[ \begin{matrix} C^T & A^T C^T & \cdots & {A^T}^{s-1}
    C^T \end{matrix} \right]$. Then we can state the
following alternative data equation:
\[
 Y_{s,N} = O_s X_N + T_{u,s}^\prime U_{s,N} + T_{e,s} E_{s,N}
\]
By this data equation, we have that,
\[
 \left[ \begin{matrix} X_N \\
     U_{s,N} \\ Y_{s,N} \end{matrix} \right] = \left[ \begin{matrix} I
     & 0 & 0 \\  0 & I & 0  \\ O_s & T_{u,s}^\prime & T_{e,s} \end{matrix} \right]\left[ \begin{matrix} X_N \\
     U_{s,N} \\ E_{s,N} \end{matrix} \right]
\]
The results follows from \eqref{posdef} and the fact that the matrix
$T_{e,s}$ is square and invertible.  \QED

Based on this result the following assumption is stipulated. 

{\bf Assumption A.2.} Consider the model~\eqref{ssmOc}, then there
exists an integer $N$ such that the compound matrix,
\[
  \left[ \begin{matrix} X_N \\
     U_{s,N} \\ Y_{s,N} \end{matrix} \right]
\]
has full row rank. 

\section{N2SID}
\label{sec4}
 \subsection{Pareto optimal Subspace Identification}

When assuming the optimal observer given, the quantity $\hat{y}(k)$ is
the minimum variance
prediction of the output and equal to $y(k) -
e(k)$. Let the Hankel matrix $\hat{Y}_{s,N}$ be defined from this sequence $\hat{y}(k)$
as we defined $Y_{s,N}$ from $y(k)$. Then the data equation \eqref{DataEq}
can be reformulated into:
\begin{equation}
 \label{DataEqPred}
\hat{Y}_{s,N} = \mathcal{O}_s X_N + T_{u,s} U_{s,N} + T_{y,s} Y_{s,N}.
\end{equation}
Let $\mathcal{T}_{p,m}$ denote the class of lower triangular block-Toeplitz
matrices with block entries $p \times m$ matrices and let $\mathcal{H}_p$ denote the class
of block-Hankel matrices with block entries of $p$ column
vectors. Then the three
key structural properties listed in Section~\ref{s_dataeq} are taken
into account in an optimization problem seeking 
a trade-off between the
following cost functions, 
\begin{equation}
 \label{NPhard}
\begin{array}{l}
\mbox{\rm rank} \Big( \Gamma_{s,N} -
{\Theta}_u U_{s,N} - {\Theta}_y  Y_{s,N} \Big) \\
\mbox{\rm and}  \quad 
\Tr \mathbb{E} [ \Big( y(k) - \gamma(k) \Big) \Big(
y(k) - \gamma(k) \Big)^T ] 
\end{array}
\end{equation}
Here $\mathbb{E}$ denotes the
expectation operator. The matrix $\Gamma_{s,N} \in
\mathcal{H}_p$ is the (block-) Hankel matrix approximating
the Hankel matrix $\hat{Y}_{s,N}$ and constructed
from the approximation of the one-step ahead prediction of the output
denoted by $\gamma(k)$ in the same way $\hat{Y}_{s,N}$ was constructed
from $\hat{y}(k)$. Further, we have the following constraints on the
matrices ${\Theta}_u \in \mathcal{T}_{p,m}$ and ${\Theta}_y \in \mathcal{T}_{p,p}$.

An optimal trade-off between the above two cost functions is called
a Pareto optimal solution. 
Moreover, the Pareto optimization problem is 
not tractable. For
that purpose we will develop in the next subsection a \emph{convex
relaxation} of the cost functions. 
This will make it possible to obtain all Pareto optimal 
solutions using scalarization. 

Before stating this convex relaxation an analysis is made on the
additional structure that can be imposed on the block-Toeplitz
matrices $\Theta_u$ and $\Theta_y$ and/or under what conditions their
block-Toeplitz structure is sufficient to find a unique solution. 

\subsection{Additional structure in the block-Toeplitz matrices
  $T_{u,s}$ and $T_{y,.s}$}

In this section we analyse the additional structure present in the block-Toeplitz matrices
  $T_{u,s}$ and $T_{y,.s}$ as well as the conditions under which the
  block-Toeplitz structure is sufficient to find the
system matrices $(\mathcal{A}_T,\mathcal{B}_T,C_T,D,K_T)$. These
conditions, 
not including the additional structural constraint highlighted in Lemma~\ref{lem2},
is summarized in Theorem~\ref{th1} of this paper. 

\begin{lemma}
 \label{lem2}
Let $s > n$, then we can partition the  block-Toeplitz matrices
  $T_{u,s}$ and $T_{y,.s}$, defined in the data
  equation~\eqref{DataEq} as,
\begin{equation}
 \label{PToep}
T_{u,s} = \left[ \begin{array}{c@{\;|\;}c} T_{u,n} & 0 \\ \hline
                   H_{u,s-n} & T_{u,s-n} \end{array} \right]
\end{equation}
and likewise for the matrix $\;T_{y,s}$. Here the matrices $H_{u,s-n}$
and $H_{y,s-n}$ can be decomposed as,
\begin{equation}
 \label{DecomH}
{\tiny \left[ \begin{array}{c@{\;|\;}c} H_{u,s-n} & H_{y,s-n} \end{array} \right] =
\left[ \begin{matrix} C \\ C \mathcal{A} \\ \vdots \\ C
    \mathcal{A}^{s-n-1} \end{matrix} \right]
\left[ \begin{array}{ccc@{\;|\:}ccc} \mathcal{A}^{n-1} \mathcal{B} &
                                                                     \cdots
         & \mathcal{B} & \mathcal{A}^{n-1} K & \cdots & K \end{array}
     \right]
}
\end{equation}
\end{lemma}

{\bf Proof:} Follows by construction. \QED

\begin{remark} \label{rem3}
Lemma~\ref{lem2} can be used to impose an additional
constraint on the block-Toeplitz matrices  $\Theta_u$ and
$\Theta_y$. If we partition these block-Toeplitz matrices conformal
their counterparts $T_{u,s}$ and $T_{y,s}$ as highlighted in
Lemma~\ref{lem2}, as follows, 
\[
 {\Theta}_{u,s} = \left[ \begin{array}{c@{\;|\;}c} {\Theta}_{u,n} & 0 \\ \hline
                   H^\Theta_{u,s-n} & {\Theta}_{u,s-n} \end{array} \right]
\]
(likewise for ${\Theta}_{y,s}$), then for the case $s \geq 2n$ we can impose the following
additional constraint,
\[
 \mbox{\rm rank} \Big( \left[ \begin{matrix} H^\Theta_{u,s-n} &
     H^\Theta_{y,s-n} \end{matrix} \right] \Big) = n
\]
\end{remark}
The additional constraint highlighted in Remark~\ref{rem3} can
be reformulated, as done e.g. in \cite{Smith14,Smith12}, as a rank
minimization constraint, that can be relaxed to a convex constraint
using the nuclear norm. However we seek to avoid imposing this additional
constraint in order to minimize the number of regularization
parameters. The basis herefore is provided in the next Theorem. 

\begin{theorem} \label{th1}
\setcounter{thm}{3}
Consider the observer in \eqref{ssmI}  with
$x(k) \in \mathbb{R}^n$ and 
consider the rank optimization problem in Eq. \eqref{NPhard} only with
$\Gamma_{s,N}$ fixed to $\hat{Y}_{s,N}$, let $s>n$ and let Assumptions
A.1 and A.2 be satisfied, Then,
\[
 \min_{\Theta_u \in  \mathcal{T}_{p,m},
  \Theta_y \in \mathcal{T}_{p,p} } \mbox{\rm rank} \Big( \hat{Y}_{s,N} -
\Theta_u U_{s,N} - \Theta_y  Y_{s,N} \Big) = n
\]
Further the arguments optimizing the above optimization problem,
denoted as $\hat{\Theta}_u, \hat{\Theta}_y$ are unique and equal to,
\[
 \hat{\Theta}_u = T_{u,s} \quad \hat{\Theta}_y = T_{y,s}
\]
with $T_{u,s}, T_{y,s}$ the true underlying block-Toeplitz matrices in
the data equation\eqref{DataEq}. 
\end{theorem}

{\bf Proof:} Let
$\delta_u \in \mathcal{T}_{p,m}, \delta_y \in \mathcal{T}_{p,p}$, then, 
\begin{eqnarray*}
 \hat{Y}_{s,N}  - \Theta_u U_{s,N} - \Theta_y  Y_{s,N} & = &
                                                                       \hat{Y}_{s,N}
                                                                       -
                                                                       (T_{u,s}+\delta_u)
                                                                       U_{s,N}
                                                                       -
  \\ & & 
                                                                       (T_{y,s}+\delta_y)
                                                                       Y_{s,N}
  \\
& = & \mathcal{O}_s X_N - \delta_u U_{s,N} - \delta_y Y_{s,N}
\end{eqnarray*}
Therefore,
\[
 \mbox{\rm rank} \Big( \hat{Y}_{s,N} -
\Theta_u U_{s,N} - \Theta_y  Y_{s,N} \Big) = 
\]
\[
\mbox{\rm rank}
\Big( \mathcal{O}_s X_N - \delta_u U_{s,N} - \delta_y Y_{s,N} \Big)
\]
Application of Sylvester's inequality \cite{VerBook} and under
Assumption A.2, we further have,
\begin{equation}
 \label{RankCond}
 \mbox{\rm rank} \Big( \hat{Y}_{s,N} -
\Theta_u U_{s,N} - \Theta_y  Y_{s,N} \Big) = \mbox{\rm rank}
\Big( \left[ \begin{matrix} \mathcal{O}_s & \delta_u &
    \delta_y \end{matrix} \right] \Big)
\end{equation}
First notice that under Assumption A.1 the rank of this matrix is $n$ for $\delta_u=0$ and
$\delta_y=0$. Since the rank$\Big( \left[ \begin{matrix} \mathcal{O}_s & \delta_u &
    \delta_y \end{matrix} \right] \Big) \geq$ rank $\Big(
\mathcal{O}_s \Big)$
for all $\delta_u, \delta_y$, we have that $n$ is the minimal value of
the rank in \eqref{RankCond}. 

It will now be shown that this minimal value of the rank, can only be
reached for both
$\delta_u$ and $\delta_y$ equal to zero. 

For that purpose, let $t = \{ t_i \in
\mathbb{R}^{p \times (m+p)} \}_{i=1}^s$
be a sequence of arbitrary matrices that 
define the lower triangular block-Toeplitz matrix $\Delta^s(t)$ as:
\[
 \Delta^s(t) = \left[ \begin{matrix}  t_1 & 0 & \cdots & 0 \\ t_2 & t_1 &
     & \vdots \\ \vdots &  & \ddots & \vdots \\ t_s & t_{s-1} & \cdots
     & t_1 \end{matrix} \right] \in \mathbb{R}^{sp \times s(m+p)}
\]
The columns of the compound matrix $\left[ \begin{matrix} \delta_u &
    \delta_y \end{matrix} \right]$ in \eqref{RankCond} can always be
permuted into a matrix of
the form $\Delta^s(t)$ and since column permutations do not change the
rank of a matrix we have that, 
\[
 \mbox{\rm rank}\Big( \left[ \begin{matrix} \mathcal{O}_s & \delta_u &
    \delta_y \end{matrix} \right] \Big) =  \mbox{\rm rank}\Big(
\left[ \begin{matrix} \mathcal{O}_s & \Delta^s(t) \end{matrix} \right] \Big)
\]
Now we show that the following condition
\[
 \mbox{\rm rank}\Big( \left[ \begin{matrix} \mathcal{O}_s &
     \Delta^s(t) \end{matrix} \right] \Big) = n
\]
implies that $\Delta^s(t)$ has to be zero. In order for the above rank
constraint to hold we need $\Delta^s(t)$ to be of the following form:
{\small \begin{equation}
 \label{KeyEq}
  \left[ \begin{matrix}
 t_1 & 0 & \cdots & 0 & 0 \\
 t_2 & t_1 &       & 0 & 0 \\ 
\vdots &   & \ddots &  &   \\
t_s & t_{s-1} & \cdots & t_2 & t_1 \end{matrix} \right] =
\mathcal{O}_s \left[ \begin{matrix} q_1 & q_2 & \cdots & q_{s-1} & q_s \end{matrix}\right]
\end{equation}}
The fact that $s > n$, we have that rank$\Big( \mathcal{O}_{s-1} \Big)
= n$ and therefore we can deduce from the first $p(s-1)$ rows of the
last $p+m$ columns in the expression \eqref{KeyEq} that,
\[
 q_s = 0 \Rightarrow t_1 = 0
\]
Using this result, and the Toeplitz structure of $\Delta^s(t)$, we
can in the same way conclude from the first $p(s-1)$ rows and from the columns
$(s-2)(m+p)+1$ to $(s-1)(m+p)$ in \eqref{KeyEq} that, 
\[
 q_{s-1} = 0 \Rightarrow t_2 = 0 \quad {\rm etc. }
\]
Hence there cannot be a $\Delta^s(t)$ with the given
Toeplitz structure that is  different from
zero such that  \\  rank$\Big(\left[ \begin{matrix}
    \mathcal{O}_s & \Delta^s(t) \end{matrix} \right]\Big) = n$. Hence
the minimal value of the rank of the matrix $\left[ \begin{matrix} \mathcal{O}_s & \delta_u &
    \delta_y \end{matrix} \right] $ in \eqref{RankCond}
w.r.t. $\delta_u, \delta_y$ yields zero value of both. This concludes
the proof. 
\QED

 \subsection{A convex relaxation}

A convex relaxation of the NP hard problem formulation in
\eqref{NPhard} will now be developed. The
original problem is reformulated in two ways. First, the 
rank operator is substituted by the nuclear norm. The nuclear
norm of a matrix $X$ denoted by $\| X \|_\star$ is defined as the
sum of the singular values of the matrix $X$. It is also known
as the trace norm, the Ky Fan norm or the Schatten
norm, \cite{Vandenberghe}. This is known to be a good approximation 
of the rank operator when it is to be minimized,   
\cite{FHB:01,Faz:02}.
Second, the minimum variance criterium is
substituted by the following sample average 
$\frac{1}{N} \sum_{k=1}^N \| y(k) - \gamma(k) \|_2^2.$ \\
By introducing a scalarization parameter  $\lambda\in[0,\infty)$, 
which can be interpreted as a 
regularization parameter,
all Pareto optimal
solutions of the convex
reformulation of the N2SID problem can be obtained by solving:
\vspace*{-1cm}
{\footnotesize 
\begin{equation}
 \label{N2SID} 
\!\!\!\!\!\!\!\!\!\!\!\!\!\!\!\!\! \left. \begin{array}{l}
 \min_{\mbox{\tiny $\Gamma_{s,N} \in \mathcal{H}_p, \Theta_{u,s} \in  \mathcal{T}_{p,m},
  \Theta_{y,s} \in \mathcal{T}_{p,p}$ }} \|  \Gamma_{s,N} -
\Theta_{u,s} U_{s,N} - \Theta_{y,s} Y_{s,N} \|_\star \\[5pt] \quad 
   + \frac{\lambda}{N} \sum_{k=1}^N \| y(k) -
   \gamma(k) \|_2^2 \end{array} \right.  .
\end{equation}
}
for all values of $\lambda\in[0,\infty)$. 

%
\begin{remark} 
\label{remOO}
The method can be extended to other related
identification problems. For example one way to consider the
identification problem of an innovation model with absence of a
measurable input, is to consider the following convex relaxed
problem formulation:
\begin{equation}
 \label{N2SIDoutputonly}
\begin{array}{l}
\min_{\Gamma_{s,N} \in \mathcal{H}_p, 
  \Theta_{y,s} \in \mathcal{T}_{p,p} } \|  \Gamma_{s,N} -
\Theta_{y,s} Y_{s,N} \|_\star +  \\ 
   \frac{\lambda}{N} \sum_{k=1}^N \| y(k) -
   \gamma(k) \|_2^2. \end{array} 
\end{equation}
\end{remark}
It is well-known that the problem \eqref{N2SID} can be recast as a Semi-Definite 
Programming (SDP) problem, \cite{FHB:01,Faz:02}, 
and hence it can be solved in polynomial time with standard SDP
solvers. The reformulation, however, introduces additional matrix variables 
of dimension $N\times N$, 
unless the problem is not further approximated using
randomization techniques as in
\cite{DBLP:journals/corr/VerhaegenH14}. 
In section~\ref{sec:ADMM} we will 
present an alternative exact 
method using ADMM inspired by its successful application in 
\cite{liu+han+van13}.

 \subsection{Calculation of the system matrices}
\label{s_calc}

The convex-optimization problem \eqref{N2SID} yields the estimates
of the quantities $\Gamma_{s,N}, \Theta_{u,s}$ and $ \Theta_{y,s} $. Since the
outcome depends on the regularization parameter $\lambda$, let us denote
these estimates as $\hat{\Gamma}_{s,N}(\lambda), \hat{\Theta}_{u,s}(\lambda)$ and
  $\hat{\Theta}_{y,s}(\lambda)$ respectively. The determination of the system matrices
  starts with an SVD of the ``low rank'' approximated matrix as follows:
\begin{eqnarray}
 \label{svd}
 \hat{\Gamma}_{s,N}(\lambda) - \hat{\Theta}_{u,s}(\lambda) U_{s,N} - \hat{\Theta}_{y,s}(\lambda) Y_{s,N}
 & = &  \nonumber \\
 & & \hspace*{-4.5cm}~\left[ \begin{array}{c@{\;|\;}c}  U_{\hat{n}}(\lambda) & 
 \star \end{array} \right] \left[ \begin{array}{c@{\;|\;}c} \Sigma_{\hat{n}}(\lambda) & 0 \\ \hline  0 &
     \star \end{array} \right] 
    \left[ \begin{array}{c} V_{\hat{n}}^T(\lambda) \\ \hline 
     \star \end{array} \right] 
\end{eqnarray}
where $\hat{n}$ is an integer denoting the $\hat{n}$ largest singular
values and the notation $\star$ denotes a compatible matrix not of
interest here. The selection of $\hat{n}$ is outlined in the
algorithmic description given next. 

The algorithm requires in addition to the input-output data
sequences the user to specify the parameter $s$ to  fix the number
of block rows in the block-Hankel matrices $U_{s,N}$ and $Y_{s,N}$ and
an interval for the parameter $\lambda$ denoted by $\Lambda =
[\lambda_{\rm min},\lambda_{\rm max}]$. 
As for the implementation described in \cite{liu+han+van13},  which we
will refer to as {\tt WNNopt}, the identification data set could be
partitioned in two parts. The first part is referred to as the {\tt
  ide-1} part of the identification data set and the remaining part of
the identification data set is referred to as the {\tt ide-2}
part. This splitting of the data set was recommended in
\cite{liu+han+van13} to avoid overfitting. In the \emph{N2SID} algorithm three variants can be substituted in the
algorithmic block  '{\tt compute
  $\mathcal{M}_j(\lambda)$}' for $j=1,2,3$. This algorithmic block
performs the actual calculation of the one-step ahead predictor and the
three variants are summarized after the description of the core part
of \emph{N2SID}.

{\tt N2SID} algorithm: \\[3pt]
{\tt 
Grid the interval $\Lambda = [\lambda_{\rm min},\lambda_{\rm max}]$ in
$N$ different points, e.g. using the Matlab \\ notation $\Lambda = {\tt logspace}\Big(
\log(\lambda_{\rm min}), \log(\lambda_{\rm max}),L \Big)$
}

{\tt
for i=1:L,  }
\begin{description}
 \item[{\tt Solve}] \eqref{N2SID} for $\lambda = \Lambda(i)$ and
data set ide-1. 
 \item[{\tt Compute}] the SVD as in \eqref{svd} for $\lambda =
\Lambda(i)$. 
 \item[{\tt Select}] Select the model order $\hat{n}$ form the singular
   values in \eqref{svd}.
 This can be done manually by the user or automatically. Such automatic selection can 
be done as in the {\tt N4SID} implementation in 
\cite{Ljungtb:07} as
highlighted  
in \cite{liu+han+van13}: order the 
singular values in \eqref{svd}
in descending order, then select that index of the singular
value  
that in logarithm is closest to the 
logarithmic mean of the maximum and 
minimum singular values in \eqref{svd}. 
 \item[{\tt Compute}] system matrices $\{ \hat{\mathcal A}_T, \hat{\mathcal
  B}_T, \hat{\mathcal C}_T, \hat{\mathcal D}, \hat{K}_T \}$ according to the procedure 'Compute 
 $\mathcal{M}_j(\lambda)$' for
$j=1,2,3$ and $\lambda =
\Lambda(i)$.
 \item[ {\tt Using}] the estimated system matrices $\{ \hat{\mathcal A}_T, \hat{\mathcal
  B}_T, \hat{\mathcal C}_T, \hat{\mathcal D}\}$, and the validation data in ide-2, compute the simulated output
$\hat{y}(k,\lambda)$ as,
\begin{eqnarray}
 \label{SimO}
 \hat{x}_T(k+1) & = & \hat{\mathcal A}_T \hat{x}_T(k) + \hat{\mathcal B}_T
 u(k) \nonumber \\ 
 \hat{y}(k,\lambda) & = & \hat{C}_T \hat{x}_T(k) + \hat{\mathcal D} u(k) 
\end{eqnarray}  and evaluate the cost function,
\[
 J(\lambda) = \sum_{i=1}^N \| y(k) - \hat{y}(k,\lambda) \|_2^2
\]
\end{description}
{\tt 
end \\[3pt]
Select $\mathcal{M}_j(\lambda_{\rm opt})$ with $\lambda_{\rm opt}$
given as:
\[
 \lambda_{\rm opt} = \min_{\lambda \in \Lambda} J(\lambda)
\]
}

The subsequent three ways to compute the model are summarized as:

\noindent
{\tt Compute $\mathcal{M}_1(\lambda)$:}
 \begin{description}
  \item[STEP 1:] 
From the SVD in \eqref{svd}, and the selected model order $\hat{n}$, 
the pair $\hat{\mathcal A}_T, \hat{C}_T$ is derived from the
    matrix $U_{\hat{n}}$ as done in classical SID methods by
    considering $U_{\hat{n}}$ to be an approximation of the extended
    observability matrix ${\mathcal O}_s$, see e.g. \cite{VerBook}.
 \item[STEP 2:] With $U_{\hat{n}}$ and the estimated matrix $ T^e_{y,s}$ we
   exploit that the latter matrix approximates the block-Toeplitz
   matrix $ T_{y,s}$ to estimate the observer gain $\hat{K}_T$ via the
   solution of a standard linear least squares problem. 
This is seen as follows. Let us assume the block Toeplitz matrix
$T_{y,s}$ be given and denoted explicitly as,
\[
 T_{y,s} =  \left[ \begin{matrix} 0 & 0 &  \cdots & 0 & 0 \\ C
    K & 0 &   & 0 & 0 \\ C \mathcal{A} K & C K &   & 0 & 0  \\ \vdots &
     &     & \ddots &  \\
C \mathcal{A}^{s-2} K &   & \cdots  & CK & 0    \end{matrix} \right] 
\]
If we know the matrix $\mathcal{O}_s$, we can write the following
set of equations,
\[
 \mathcal{O}_s(1:(s-1)p,:) K = T_{y,s}(p+1:ps,1:p) 
\] 
Let us now denote the first $(s-1)p$
rows of the matrix $U_{\hat{n}}(\lambda_i)$  by
$\mathcal{\hat{O}}_{s-1,T}$ 
and let us denote the submatrix of
the matrix $\hat{\Theta}_{y,s}(\lambda_i)$  from rows $p+1$ to row
$ps$ and from column $1$ to $p$ by $\hat{T}_{y,s}(p+1:ps,1:p)$, then
we can estimate $K_T$ from:
\begin{equation}
 \label{estK}
\min_{K_T} \|  \mathcal{\hat{O}}_{s-1,T} K_T -  \hat{T}_{y,s}(p+1:ps,1:p)\|^2
\end{equation} 
This
   estimate of the observer gain is used to estimate the system matrix
   $A_T$ as:
\begin{equation}
 \label{Acomp}
 \hat{A}_T = \hat{\mathcal A}_T + \hat{K}_T \hat{C}_T
\end{equation}
 \item[STEP 3:] Let the approximation of the observer
   be denoted as:
\begin{eqnarray}
 \label{ObsE}
 \hat{x}_T(k+1) & = & \hat{\mathcal A}_T \hat{x}_T(k) + \hat{\mathcal B}_T
 u(k) + \hat{K}_T y(k) \nonumber \\ 
 \hat{y}(k) & = & \hat{C}_T \hat{x}_T(k) + \hat{\mathcal D} u(k) 
\end{eqnarray}
Then the estimation of the
pair $\hat{\mathcal B}_T, \hat{D}$ and the initial conditions of the above
observer can again be done via a linear least squares problem as
outlined in \cite{VerBook} by minimizing the RMS value of the prediction
error $y(k) - \hat{y}(k)$ determined from the identification data in {\tt ide-1}.
The estimated input matrix $\hat{B}_T$ is then determined as:
\begin{equation}
 \label{Bcomp}
 \hat{B}_T = \hat{\mathcal B}_T + \hat{K}_T D 
\end{equation}
 \end{description}

\noindent
{\tt Compute $\mathcal{M}_2(\lambda)$:}
 \begin{description}
  \item[STEP 1:] as in {\em Compute $\mathcal{M}_1(\lambda)$.}
  \item[STEP 2:] Derive an estimate of the state sequence of the
    observer \eqref{ObsE} from the SVD \eqref{svd}, where for the sake of compactness again
    the system symbol $\hat{x}_T(k)$ will be used, 
\[
 \left[ \begin{matrix} \hat{x}_T(1) & \hat{x}_T(2) & \cdots &
     \hat{x}_T(N-s+1) \end{matrix} \right] \approx V_{\hat{n}}^T( \lambda)
\]
Using the singular values this approximation could also be scaled
as $\sqrt{\Sigma_{\hat{n}} ( \lambda) } V_{\hat{n}}^T(
\lambda) $. 
   \item[STEP 3:]  Knowledge of the estimated state sequence of the
     observer \eqref{ObsE} turns the estimation of the system matrices
     $\hat{\mathcal A}_T,  \hat{\mathcal B}_T, \hat{C}_T,\hat{\mathcal
       D},\hat{K}_T$ 
      and the observer inititial conditions into linear
     least squares problem. The estimated pair $(\hat{A}_T,\hat{B}_T)$
     can be computed from this
     quintuple as outlined in \eqref{Acomp} and
     \eqref{Bcomp}, respectively. 
 \end{description}

\noindent
{\tt Compute $\mathcal{M}_3(\lambda)$:}

\noindent
 \begin{description}
  \item[STEP 1 and 2:] as in {\em Compute $\mathcal{M}_1(\lambda)$.}
  \item[STEP 3:] With $U_{\hat{n}}$ and the estimated Markov
    parameters in $T^e_{u,s}$ we could similarly to estimating the
    Kalman gain, also estimate the pair $\hat{\mathcal B}_T, \hat{D}$ via a linear
    least squares problem. The matrix
    $\hat{B}_T$ can be estimated from $\hat{\mathcal B}_T$ as outlined in
    \eqref{Bcomp}, 
 \end{description}
In the experiments reported in Section~\ref{s_sim} use will be made of
{\emph N2SID Algorithm} with the model computation block {\tt Compute
  $\mathcal{M}_1(\lambda)$}. It turned out that the
{\tt N2SID} algorithm is much less sensitive to overparametrization
compared as compared to {\tt WNNopt}. For that reason we will
use the whole identification data set in 
all steps of the
{\tt N2SID} algorithm for the experiments reported in
Section~\ref{s_sim}, i.e. {\tt ide-1} and {\tt ide-2} are identical
and equal to the identification data set. 

\section{ADMM}
\label{sec:ADMM}
The problem we like to solve is exactly of the form in (20) in 
\cite{liu+han+van13}, i.e.
\begin{equation}
\min_x\| \mathcal A(x)+A_0\|_\star +\frac{1}{2}(x-a)^TH(x-a)  \label{ADMM} 
\end{equation}
for some linear operator $\mathcal A(x)$ and some positive semidefinite 
matrix $H$. In the above mentioned reference the linear operator is a Hankel
matrix operator, and this structure is used to tailor the ADMM code to run 
efficiently. Essentially the key is to be able to compute the coefficient 
matrix related to the normal equations of the linear operator in an
efficient way using FFT. This matrix $M$
is defined via 
$$\mathcal A_\adj(\mathcal A(x))=Mx,\;\forall x$$
where $\mathcal A_\adj(\cdot)$ is the adjoint operator of $\mathcal A(\cdot)$.
Similar techniques 
have been used for Toeplitz operators in \cite{roh+van06}, and are closely
related to techniques for exploiting Toeplitz structure in linear systems
of equations, \cite{gol+loa96}. For our problem the linear operator 
consists of a sum of Hankel and Toeplitz operators, and we will show how FFT
techniques can be used also for this operator. For more details on the 
ADMM algorithm see the appendix. 
\subsection{Circulant, Toeplitz and Hankel Matrices}
We define the circulant matrix operator 
$\mathcal C^{n}:\Re^n\rightarrow \Re^{n\times n}$ 
of a vector $x\in\Re^n$ via
\begin{equation}
\mathcal C^n(x)=
\begin{bmatrix}
x_1&x_{n}&\cdots&x_{3}&x_2\\
x_2&x_1&x_n& &x_3\\
\vdots&x_2&x_1&\ddots&\vdots\\
x_{n-1}& &\ddots&\ddots&x_n\\
x_n&x_{n-1}&\cdots&x_2&x_1
\end{bmatrix}.
\end{equation}
We also define the Hankel matrix operator 
$\mathcal H^{(m,n)}:\Re^{m+n-1}\rightarrow \Re^{m\times n}$ 
of a vector $x\in\Re^{m+n-1}$ via
\begin{equation}
\mathcal H^{(m,n)}(x)=
\begin{bmatrix}
x_1&x_{2}&\cdots&x_n\\
x_2&\reflectbox{$\ddots$}& &\vdots\\
\vdots& & &\vdots\\
x_m&\cdots&\cdots&x_{m+n-1}
\end{bmatrix}.
\end{equation}
For a vector$x\in\Re^{m+n-1}$ it holds that $\mathcal H^{(m,n)}(x)=
\mathcal C^{m+n-1}_{n:n+m-1,n:-1:1}(x)$, i.e. the Hankel operator is the 
lower left corner of the circulant operator where the columns are
taken in reverse order. We also define the Toeplitz operator 
$\mathcal T^{n}:\Re^{2n-1}\rightarrow \Re^{n\times n}$ 
of a vector $x\in\Re^{2n-1}$ via
\begin{equation}
\mathcal T^{n}(x)=
\begin{bmatrix}
x_n&x_{n-1}&\cdots&x_1\\
x_{n+1}&\ddots& &\vdots\\
\vdots& &\ddots &\vdots\\
x_{2n-1}&\cdots&\cdots&x_{n}
\end{bmatrix}.
\end{equation}
We realize that $\mathcal T^{n}(x)=\mathcal H^{(n,n)}_{:,n:-1:1}(x)$, i.e. a Toplitz 
operator can be obtained from a square Hankel operator by taking the columns 
in reverse order. We are finally interested in upper triangular
Toeplitz operators with and without zeros on the diagonal, 
and we remark that these are easily obtained from the normal Toeplitz operator
by replacing $x$ with $\begin{bmatrix}x^T & 0\end{bmatrix}^T$, which will be 
upper triangular with a non-zero diagonal if $x\in\Re^n$ and with a zero
diagonal if $x\in\Re^{n-1}$. 
\subsection{The Fourier Transform and Hankel Matrices}
It is well-known, \cite{gol+loa96},
that if we let $\mathcal F^n\in\Co^{n\times n}$ 
be the discrete Fourier transform
matrix of dimension $n$, then the circulant matrix can be 
expressed as
\begin{equation}
\mathcal C^n(x)=\frac{1}{n}(\mathcal F^n)^H\diag(\mathcal F^nx)\mathcal F^n.
\end{equation}
From this we immediately obtain that the Hankel matrix can be expressed
as 
\begin{equation}
\mathcal H^{(m,n)}(x)=\frac{1}{N}H^H\diag(\mathcal F^Nx)G
\end{equation}
where $N=n+m-1$, $F=\mathcal F^N$, $G=F_{:,n:-1:1}$ and $H=F_{:,n:n+m-1}$.
This expression will make it easy for us to represent the adjoint of the 
Hankel operator. It is straight forward to verify that the adjoint 
$\mathcal H_\adj^{(m,n)}(Z):\Re^{m\times n}\rightarrow\Re^{n+m-1}$ is given
by
$$\mathcal H_\adj^{(m,n)}(Z)=\frac{1}{N}F^H\diag(HZG^H).$$
Notice that we are abusing the operator $\diag(\cdot)$. In case the argument
is a vector the operator produces a diagonal matrix with the vector on the 
diagonal, and in case the argument is a square matrix, the operator produces
a vector with the components equal to the diagonal of the matrix. 
\subsection{The Linear Operator $\mathcal A$}
We will now present the linear operator that we are interested in for the 
SISO case:
$\mathcal A:\Re^N\times\Re^s\times\Re^{s-1}\rightarrow \Re^{s\times n}$, where
{\tiny 
\[ \mathcal A(x)=\mathcal H^{(s,n)}(\hat{y})+
\mathcal T^s\left(\begin{bmatrix}v_{s:-1:1}\\0\end{bmatrix}\right)^TV
+\mathcal
T^s\left(\begin{bmatrix}w_{s-1:-1:1}\\0\end{bmatrix}\right)^TW
\]
}
%
where $n=N-s+1$, 
$x=(\hat y,v,w)$ with $\hat y\in\Re^N$, $v\in\Re^s$, $w\in\Re^{s-1}$, 
$V\in\Re^{s\times N}$, and $W\in\Re^{s\times N}$. By taking  $V = -U_{s,N}$ and 
$W=-Y_{s,N}$ we obtain the linear operator for N2SID. Then, $v$ and $w$ 
are the first columns of the Toeplitz matrices $T_{u,s}$ and $T_{y,s}$, 
respectively.
We can express $\mathcal A(x)$ in
terms of Hankel operators as:
%
{\tiny 
\[
\!\!\!\!\!\!\!\!\! \mathcal A(x)=\mathcal H^{(s,n)}(\hat y)+
\mathcal H^{(s,s)}_{:,s:-1:1}\left(\begin{bmatrix}v_{s:-1:1}\\0\end{bmatrix}\right)^TV
+\mathcal H^{(s,s)}_{:,s-1:-1:1}\left(\begin{bmatrix}w_{s-1:-1:1}\\0\end{bmatrix}\right)^TW.
\]
}
%
The adjoint of this operator 
can be expressed in terms of the adjoint of the Hankel operator as
$$\mathcal A_\adj(Z)=\begin{bmatrix}\mathcal H_\adj^{(s,n)}(Z)\\
\mathcal H^{(s,s)}_{\adj,s:-1:1}(VZ_{s:-1:1,:}^T)\\
\mathcal H^{(s,s)}_{\adj,s-1:-1:1}(WZ_{s:-1:1,:}^T)\end{bmatrix}$$
\subsection{Forming the Coefficient Matrix}
A key matrix in the ADDM algorithm is the matrix $M$ defined via
$$\mathcal A_\adj(\mathcal A(x))=Mx,\quad \forall x.$$
We will now show how this matrix can be formed efficiently using the Fast
Fourier Transform (FFT). We partition the matrix as
$$M=\begin{bmatrix}M_{11}&M_{12}&M_{13}\\M_{12}^T&M_{22}&M_{23}\\M_{13}^T&M_{23}^T&
M_{33}\end{bmatrix}$$
where the partition is done to conform with the partition $x=(y,v,w)$. It is
then clear that $M_{11}$ is defined via
$$\mathcal H^{(s,n)}_\adj(\mathcal H^{(s,n)}(y))=M_{11}y,\quad\forall y$$
The left hand side can be expressed as
$$\frac{1}{N^2}F^H\diag\left(HH^H\diag\left(Fy\right)
GG^H\right).$$
From the identity
$$\diag\left(A\;\diag\left(x\right)
B\right)=(A\odot B^T)x$$
where $\odot$ denotes the Hadamard product of matrices, it follows that 
$$M_{11}=
\frac{1}{N^2}F^H\left(\left(HH^H\right)\odot\left(\overline{GG^H}
\right)\right)F$$
The efficient way to form $M_{11}$ is to first compute $F$,$G$ and $H$ using an
FFT algorithm, and then to form the matrix 
$$X=\left(HH^H\right)\odot\left(\overline{GG^H}\right).$$
After this one should apply the inverse FFT algorithm to $X^T$, and then 
to the transpose of the resulting matrix once more the inverse FFT algorithm. 

The expressions for the other blocks of the matrix $M$ can be derived in 
a similar way, and they are given by:
\begin{eqnarray*}
M_{12}&=&\frac{1}{N^2}F^H\left(\left(H_{:,s:-1:1}H^H\right)
\odot\left(\overline{GV^TG_{:,1:s}^H}\right)\right)F_{:,s:-1:1}\\
M_{13}&=&\frac{1}{N^2}F^H\left(\left(H_{:,s:-1:1}H^H\right)
\odot\left(\overline{GW^TG_{:,1:s}^H}\right)\right)F_{:,s-1:-1:1}\\
M_{22}&=&\frac{1}{\kappa^2}F_{:,s:-1:1}^H\left(\left(HVV^HH^H\right)
\odot\left(\overline{GG^H}\right)\right)F_{:,s:-1:1}\\
M_{23}&=&\frac{1}{\kappa^2}F_{:,s:-1:1}^H\left(\left(HVW^HH^H\right)
\odot\left(\overline{GG^H}\right)\right)F_{:,s-1:-1:1}\\
M_{33}&=&\frac{1}{\kappa^2}F_{:,s-1:-1:1}^H\left(\left(HWW^HH^H\right)
\odot\left(\overline{GG^H}\right)\right)F_{:,s-1:-1:1}.
\end{eqnarray*}
Notice that for the last three blocks the matrices $F$, $G$, and $H$ are 
defined via a discrete
Fourier transform matrix of order $\kappa=2s-1$. 
\subsection{MIMO Systems}
So far we have only discussed SISO systems. For a general $p\times m$ system
we may write the linear operator 
$\mathcal A:\Re^{pN}\times\Re^{pms}\times\Re^{pp(s-1)}\rightarrow \Re^{ps\times n}$ 
as:
$$\mathcal A(x)=\sum_{i=1}^p \mathcal A_{i}(x_{i})\otimes e_{i}$$
where 
\begin{eqnarray*}
\mathcal A_i(x_i)&=\mathcal H^{(s,n)}(\hat y_i)+\sum_{j=1}^m
\mathcal H^{(s,s)}_{:,s:-1:1}\left(\begin{bmatrix}v^{i,j}_{s:-1:1}\\0\end{bmatrix}\right)^TV_j\\
&+\sum_{j=1}^p\mathcal H^{(s,s)}_{:,s-1:-1:1}\left(\begin{bmatrix}w^{i,j}_{s-1:-1:1}\\0\end{bmatrix}\right)^TW_j
\end{eqnarray*}
where $V_j=-U_{s,N}^j$ and $W_j=-Y_{s,N}^j$ are Hankel matrices
defined from $u_j$ and $y_j$, i.e.
from the $j$th inputs and outputs, respectively, and where $e_i$ is the $i$th
basis vector. Hence we may interpret 
each term $\mathcal A_{i}(x_{i})$ as defining a MISO system in the sense that
each predicted output can be written as a linear combination of all the 
intputs and outputs. If we write the adjoint variable $Z$ in a similar way
as $Z=\sum_{i=1}^p Z_i\otimes e_{i}$, it follows that the adjoint operator
is given by $\mathcal A_\adj(Z)=(\mathcal A_{\adj,1}(Z_1),\ldots,
\mathcal A_{\adj,p}(Z_p))$. Hence the matrix $M$ will now be blockdiagonal
with blocks defined from the identity
$$\mathcal A_{\adj,i}(\mathcal A_i(x_i))=M_ix_i,\quad \forall x_i,\;i=1,\ldots,p.$$
It is not difficult to realize that the operators $\mathcal A_{\adj,i}(Z_i)$ 
will be given by
$$\mathcal A_{\adj,i}(Z_i)=\begin{bmatrix}\mathcal H_\adj^{(s,n)}(Z_i)\\
\mathcal H^{(s,s)}_{\adj,s:-1:1}(V_1Z_{i;s:-1:1,:}^T)\\
\vdots\\
\mathcal H^{(s,s)}_{\adj,s:-1:1}(V_mZ_{i;s:-1:1,:}^T)\\
\mathcal H^{(s,s)}_{\adj,s-1:-1:1}(W_1Z_{i;s:-1:1,:}^T)\\
\vdots\\
\mathcal H^{(s,s)}_{\adj,s-1:-1:1}(W_pZ_{i;s:-1:1,:}^T)\end{bmatrix}.$$
Hence each of the blocks $M_i$
will have a similar structure as the $M$ matrix for the SISO system. However,
the sub-blocks $M_{12}$, $M_{13}$, $M_{22}$, $M_{23}$ and $M_{33}$ will have 
sub-blocks themselves reflecting that fact that there are several inputs and
outputs. $M_{11}$ will be the same as $M_1$ for the SISO case for all $i$. Below
are formulas given for sub-blocks of each of the other matrices
\begin{eqnarray*}
M_{12j}&=&\frac{1}{N^2}F^H\left(\left(H_{:,s:-1:1}H^H\right)
\odot\left(\overline{GV_j^TG_{:,1:s}^H}\right)\right)F_{:,s:-1:1}\\
M_{13j}&=&\frac{1}{N^2}F^H\left(\left(H_{:,s:-1:1}H^H\right)
\odot\left(\overline{GW_j^TG_{:,1:s}^H}\right)\right)F_{:,s-1:-1:1}\\
M_{22jk}&=&\frac{1}{\kappa^2}F_{:,s:-1:1}^H\left(\left(HV_jV_k^HH^H\right)
\odot\left(\overline{GG^H}\right)\right)F_{:,s:-1:1}\\
M_{23jk}&=&\frac{1}{\kappa^2}F_{:,s:-1:1}^H\left(\left(HV_jW_k^HH^H\right)
\odot\left(\overline{GG^H}\right)\right)F_{:,s-1:-1:1}\\
M_{33jk}&=&\frac{1}{\kappa^2}F_{:,s-1:-1:1}^H\left(\left(HW_jW_k^HH^H\right)
\odot\left(\overline{GG^H}\right)\right)F_{:,s-1:-1:1}.
\end{eqnarray*}
It is interesting to notice that these formulas do not depend on index $i$. 
This means that all $M_i$ are the same.
\section{Validation Study}
 \label{s_sim}
In this section we report results on numerical experiments using
real-life data sets. We will make use of some representative data sets
from the DaISy collection, \cite{DDDF:97}. For preliminary
test with the new N2SID mehod based on academic examples, we refer to
\cite{DBLP:journals/corr/VerhaegenH14}.  

The numerical results reported in Subsection~\ref{val_res}
were performed with 
Matlab.  The implementations have been carried out in
MATLAB R2013b running on an Intel Core i7 CPU M 250 2 GHz
with 8 GB of RAM.

\subsection{Data selection and pre-processing}

From the DaISy collection, \cite{DDDF:97}, five representative data sets
were selected. These sets contain SISO, SIMO, MISO and MIMO
systems. Information about the selected data sets is provided in
Table~\ref{tab:DAISYdescr}.  
\begin{table*}[htbp]
\caption{Five benchmark problems from the DaISy collection,
  \cite{DDDF:97}; $N_{\rm tot}$ is the total number of data samples
  available}
\label{tab:DAISYdescr}
\begin{center}
\begin{tabular}{r|c|l|c|c|c}
Nr & Data set &Description & Inputs & Outputs & $N_{\rm tot}$\\
\hline
1&96-007&CD player arm & 2 & 2 & 2048 \\
2&98-002&Continuous stirring tank reactor & 1 & 2 & 7500 \\
3&96--006&Hair dryer & 1 & 1 & 1000 \\
4&97-002&Steam heat exchanger& 1& 1 & 4000 \\
5&96-011&Heat flow density&2 & 1 & 1680\\
\end{tabular}
\end{center}
\end{table*}
In order to evaluate the performances for small length data sets, data
sets of increasing length are considered. The data length is indicated
by $N_{\rm ide}$  in Table~\ref{tab:DAISYlength} for each data set of
Table~\ref{tab:DAISYdescr} . To test the sensitivity of the
identification mehods with respect to the length of the identification data set,
$N_{\rm ide}$ is increased from a small number, as compared to the total
number of samples available, in a way as indicated in Table~\ref{tab:DAISYlength}
\begin{table*}[htbp]
\caption{The increasing length $N_{\rm ide}$ of the data sets used for
  system identification starting with the sample index $del$; $N_{\rm
    val}$ indicates the length of the validation data set starting
  with the sample index $max(N_{\rm ide}) +1$. }
\label{tab:DAISYlength}
\begin{center}
\begin{tabular}{r|| l |l |l |l |l | l | l| l| l| l|| l ||  l}
Nr & \multicolumn{10}{|c||}{$N_{\rm ide}$} & $del$ & $N_{\rm val}$ \\
\hline
1& 80  & 120  & 150 &  175  &  200  & 300  & 400 &  500   & 600 &  &
120 & 500 \\
2& 100 &  150 &  200 &  300  & 400 &   500 &  600 &  700 &  800 &  &
200 & 1500 \\
3 & 80 &  100  & 120 &  140  &  160  &  180 &   200 &   250  & 300  &
400 & 120 & 600  \\
4 & 150   &  200    &  300 &   500  & 750  &  1000 & 1250 & 1500 &
1750 &  & 200 & 1500 \\
5 & 175 &   200 &  250  & 300 &   350  & 400 &  450 &  500  & 550  &
600 & 200 & 1000 
\end{tabular}
\end{center}
\end{table*}
From each identification and validation data set the offset is
removed. Data set~2 from the continuous stirred tank reactor is
scaled in such a way that both outputs have about the same numerical
range. This is achieved by scaling the detrended versions of these
outputs such that the maximum value of each output equals $1$. 

Since many of the data sets contain
poorly excited data at their beginning, the first $del$ samples are
discarded from each data set. The actual value of $del$ for each data
set is listed in Table~\ref{tab:DAISYlength}. Finally each identified
model is validated for each test case on the same validation data
set. These validation data sets contain the $N_{\rm val}$ samples
following the 
sample with index $\max(N_{\rm ide}) + 1$. The value of $N_{\rm val}$
is listed in
Table~\ref{tab:DAISYlength}. 

\subsection{Compared Identification methods}
 \label{s_methods}
Three SID methods are compared in the tests. Their key user
selection parameters are listed in Table~\ref{tab:Methods}.
\begin{table*}[htbp]
\caption{Three SID methods and their user selection parameters
  $\frac{\lambda}{N_{\rm ide}}$ and the number of block-rows in the
  data Hankel matrices $s$.}
\label{tab:Methods}
\begin{center}
\begin{tabular}{r|| r | r | r }
Method  & $\frac{\lambda}{N_{\rm ide}}$ & $s$ & {\em Weighting} \\
\hline
{\tt N4SID} [\cite{Ljungtb:07}] &  $/$  &  15 & {\em automatic} \\
{\tt WNNopt}, [\cite{liu+han+van13} ]& $[10^{-3},10^3]$ & 15 & {\em
  CVA} \\
{\tt N2SID} {\em Algorithm} & $[10^{-1.5},10^3]$ & 15  & $/$ \\
\end{tabular}
\end{center}
\end{table*}
One of the key user selection parameters of the SID methods is the number of
block rows $s$ of the Hankel data matrices. In methods like {\tt N4SID} or
{\tt WNNopt} of Table~\ref{tab:Methods} a distinction could be made in
the number of block
rows of so-called future and past Hankel matrices. Such 
differentiation is not necessary for the 
{\tt N2SID} algorithm. Since such differentation is still an open
research problem we opted in this simulation study to take the number
of block rows of the future and past Hankel matrices in  {\tt N4SID} or
{\tt WNNopt} equal to the number of block rows in the {\tt N2SID} algorithm.
Table~\ref{tab:Methods} also lists the interval of the regularization
parameter $\lambda$ to be specified for
the Nuclear Norm based methods. 

For {\tt N4SID} we further used the default settings except that the Kalman
filter gain is not estimated, a guaranteed stable simulation model
is identified, no input delays are estimated and these are fixed to
zero, and finally no covariance estimates are determined.  The order
selection is done with {\tt N4SID} using the option {\tt 'best'},
\cite{Ljungtb:07}. This results in a similar automatic
choise as we have implemented for N2SID. 

For {\tt WNNopt} in \cite{liu+han+van13} the weighting according to the
  CVA method is used. Also here no Kalman gain is estimated and the
  input delay is set to zero. As indicated in \cite{liu+han+van13} an
  'identification' and a 'validation' data set is needed to perform the
  selection of the regularization parameter $\lambda$ in order to
  avoid overfitting. In this paper both data sets are retrieved from
  the identification data set of length $N_{\rm ide}$ by splitting it
  into two almost equal parts differing in length by at most one sample. 

For N2SID we used the 
ADMM algorithm presented in \cite{liu+han+van13}, where we have provided our
own routines for computing $M$ as explained in section~\ref{sec:ADMM}. As explained in 
\cite{liu+han+van13} we also make use of simultaneous diagonalization of $M$ and
the positive semidefinite matrix $H$ in (\ref{ADMM}) in order not to have
to make different factorizations for each value of the regularization 
parameter $\lambda$. The maximum number of iterations in the ADMM algorithm have
been set to 200, the absolute and relative solution accuracy tolerances have been
set to $10^{-6}$, and $10^{-3}$, respectively. The parameters used to update the 
penalty parameter have been set to $\tau=2$ and $\mu=10$. We label 
our {\it N2SID Algorithm} with {\tt N2SID}). 
Also we do not split the data for {\tt N2SID}. We have also carried out
experiments when we did split the data. This resulted in most cases in
comparable results and in some cases even better results.   

The SID methods are compared with the prediction error method
{\tt PEM} of the matlab System Identification toolbox
\cite{Ljungtb:07}. Here the involvement of the user in specifying the
model structure is avoided by initializing {\tt PEM} with the
model determined by {\tt N4SID}. Therefore, the
model order of {\tt PEM} is the same as that 
determined by the {\tt N4SID} method. In this way no user selection
parameters are needed to be specified for {\tt PEM}.  This is
in agreement with the recommendation given on the {\tt PEM} help page
\url{http://nf.nci.org.au/facilities/software/Matlab/toolbox/ident/pem.html}
when identifying black-box state space models. 

\subsection{Results and Discussion}
\label{val_res}

The three SID methods in Table~\ref{tab:Methods} and the {\tt PEM}
method will be compared for the data sets in Table~\ref{tab:DAISYdescr}. The results of this
comparison are for each data set summarized in two graphs in the same
figure. The left graph of the figure displays the goodness of fit
criterium VAF. This is defined using the identified quadruple of
system matrices $[\hat{A}_T,\hat{B}_T,\hat{C}_T,\hat{D}]$ obtained
with each method to predict the output using the
validation data set. Let the predicted output be denoted by
$\hat{y}_v(k)$ for each method, and let the output measurement in the
validation data set be denoted by $y_v(k)$. Then VAF is 
defined as:
\begin{equation}
 \label{VAF}
{\rm VAF} = \Big( 1 - \frac{\frac{1}{N_{\rm val}}
  \sum_{k=1}^{N_{\rm val}} \| y_v(k) - \hat{y}_v(k)
  \|_2^2}{\frac{1}{N_{\rm val}} \| y_v(k) \|_2^2 } \Big) 100 \%
\end{equation}
The right graph of the figure displays the model complexity as
defined by the model order of the state space model. Both the goodness
of fit and estimated model order are graphed versus the length of the
identification data batch as indicated by the symbol $N_{\rm ide}$ in
Table~\ref{tab:DAISYlength}.

All these results are obtained in a similar ``automized manner'' for
fair comparison as outlined in section~\ref{s_methods}. In
order to evaluate the results additional information is retrieved from
the singular values as computed by the SID methods {\tt WNNopt} and
{\tt N2SID}. This is done in order to see possible improvements in the low
rank detection by the new SID method {\tt N2SID} over {\tt WNNopt}. For
an illustration of the potential improvement of the latter over {\tt
  N4SID} we refer to \cite{liu+han+van13}. 

\subsubsection{The CD player arm data set (\# 1 in
  Table~\ref{tab:DAISYdescr})}

\begin{figure}[h]
\includegraphics[width=\linewidth]{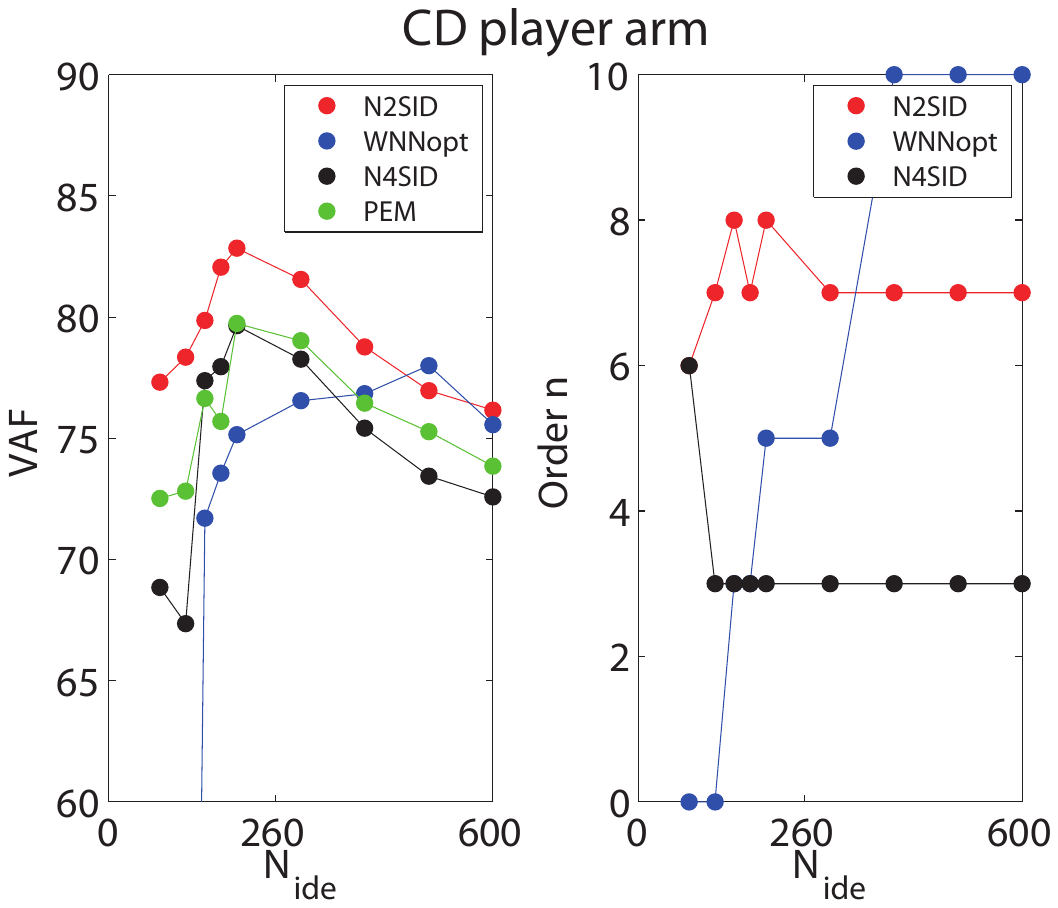}
\caption{VAF Daisy \# 1 - CD player arm.}
\label{fig:vaf_daisy1}
\end{figure}

The results are summarized in Figure~\ref{fig:vaf_daisy1}. The goodness
of fit is given on the left side of this figure and the detected order
$\hat{n}$ on the right side.  For $N_{\rm ide} \leq 400$, {\tt N2SID}
outperforms all other methods and it was always better then {\tt
  N4SID}.  {\tt PEM} is able to  improve the results of {\tt N4SID}
{\em in most cases}. Its results remain however inferior to {\tt N2SID}.
In general {\tt N2SID} detects a larger model order.
For the shortest data lengths $N_{\rm ide} = 80$ and $120$, {\tt WNNopt} was not able to
produce results since for that case the ADMM implementation broke
down. The reason being that 
the Schur form was no longer computable as it contained {\tt
  NaN} numbers. For that reason both the VAF and the order were put to 
zero. The {\tt WNNopt} determined for $150 \leq N_{\rm ide} \leq
300$ a lower model order $\hat{n}$ compared to {\tt N2SID}, but this a
the cost of a lower VAF. For $N_{\rm ide} = 150$ and $175$, the same
order as for {\tt N4SID} was detected, however with worser VAF as compared
to both {\tt N4SID} and {\tt PEM}. For  $N_{\rm ide} \geq
400$ the limit set on the model order, which was $10$ in all
experiments, was selected by {\tt WNNopt}, sometimes but  not always
yielding a better VAF.

\begin{figure}
\includegraphics[width=\linewidth]{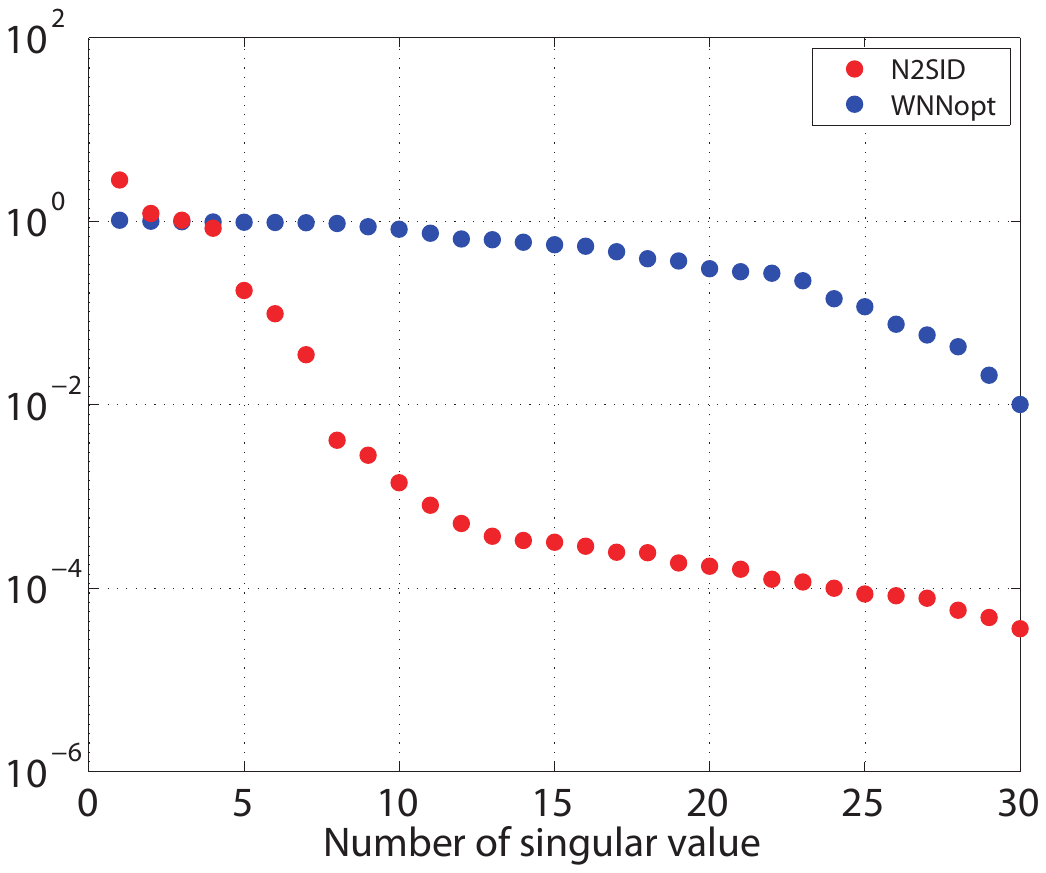}
\caption{Singular values Daisy \# 1 - CD player arm.}
\label{fig:sv_daisy1}
\end{figure}

The efffect of the use of instrumental variables and the splitting of
the identification data set to avoid overfitting on the order
selection is clear from the singular values of {\tt N2SID} and {\tt WNNopt}
 given in  Figure~\ref{fig:sv_daisy1} for  $N_{\rm ide}  = 600$. 
This plot visually supports the selection of a 7-th order model by {\tt
  N2SID} and it also explains why {\tt WNNopt} selects a larger model
order. One possible explanation is that the instrumental variables and 
projections have ``projected away'' crucial information in the data. 

\subsubsection{The Continuous stirred Tank Reactor data set (\# 2 in
  Table~\ref{tab:DAISYdescr})}

\begin{figure}[h]
\includegraphics[width=\linewidth]{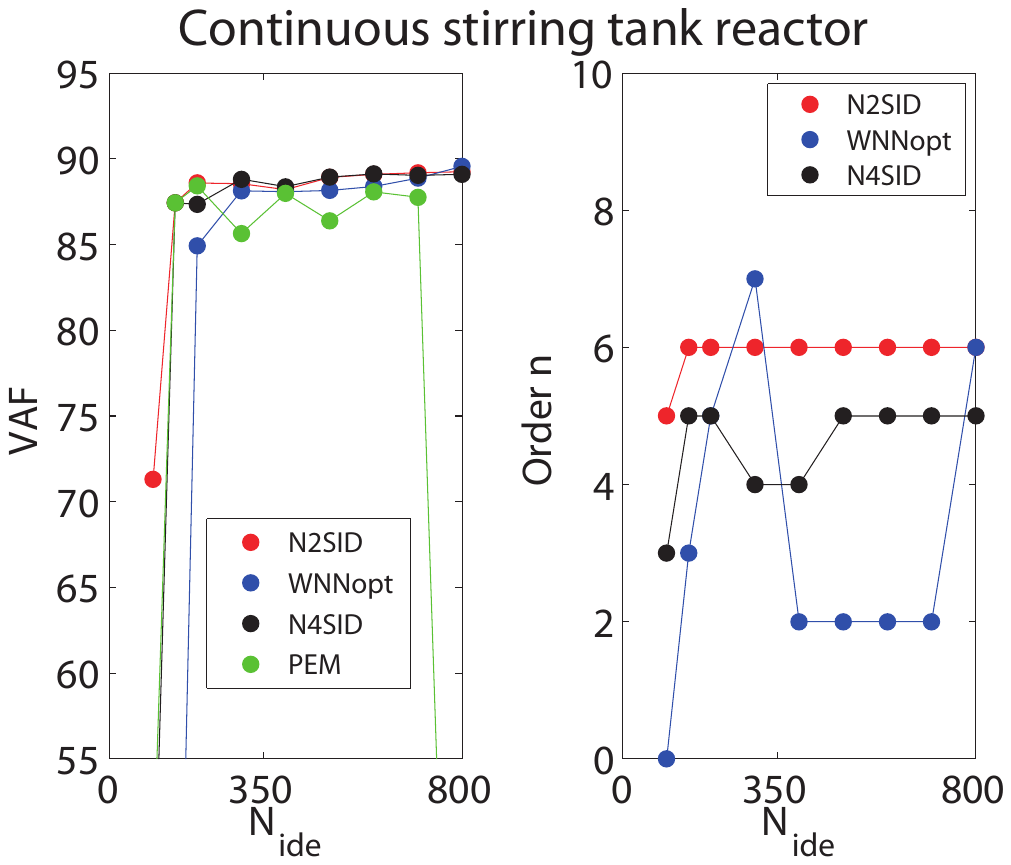}
\caption{VAF Daisy \# 2 - Continuous Stirred Tank Reactor.}
\label{fig:vaf_daisy2}
\end{figure}

The goodness of fit parameter VAF and the
estimated model order $\hat{n}$ are plotted in
Figure~\ref{fig:vaf_daisy2} in the left and right graphs, respectively. 
For $N_{\rm ide} = 100$ and $150$ {\tt WNNopt} was not able to provide
numerical results. For that reason the corresponding
VAF values are again fixed to zero.  {\tt PEM} resulted in bad VAF results for $N_{\rm
  ide} = 100$, probably as a consequence of bad initialization
from {\tt N4SID}. However, also for $N_{\rm ide} = 800$ {\tt PEM} had 
severly degraded results despite the fact that {\tt N4SID} provided a model
of comparable quality with the other SID methods. 

\begin{figure}
\includegraphics[width=\linewidth]{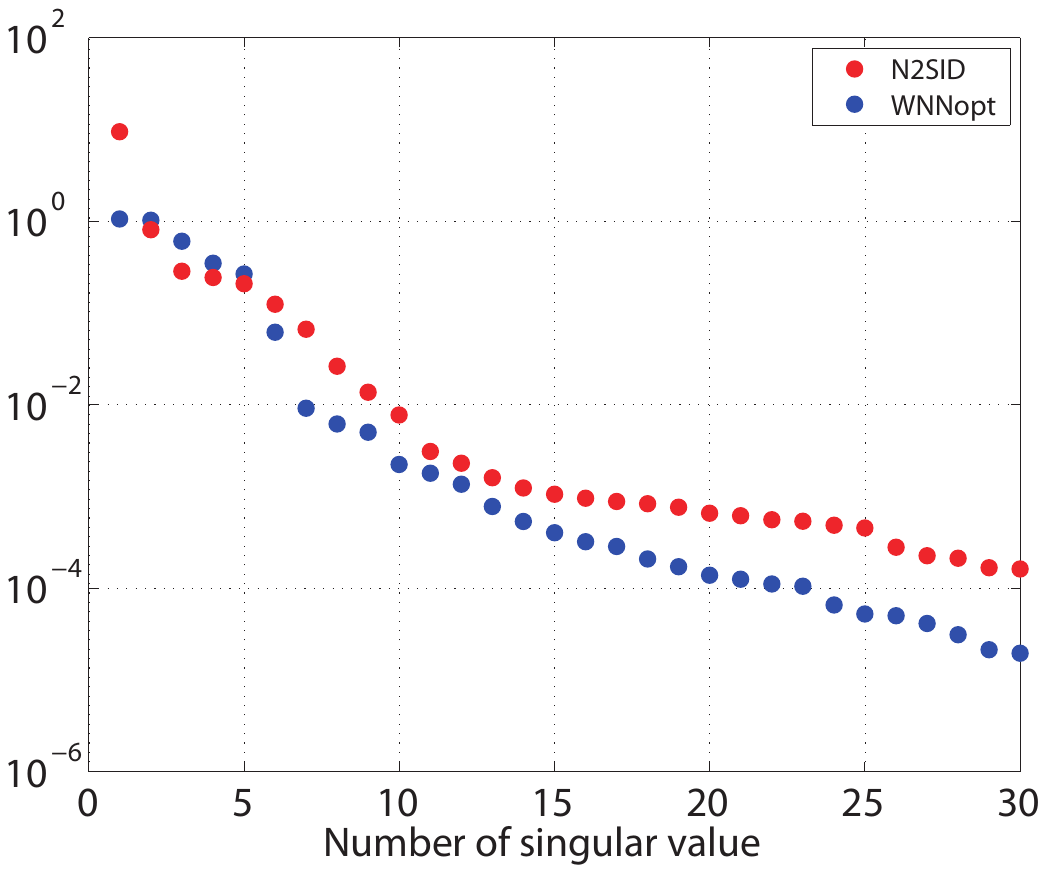}
\caption{Singular values Daisy \# 2 - Continuous Stirred Tank Reactor.}
\label{fig:sv_daisy2}
\end{figure}

The singular values in Figure~\ref{fig:sv_daisy2} indicate that for
$N_{\rm ide} = 800$ both {\tt N2SID} and {\tt WNNopt} have the same
order estimate $\hat{n}$. There is a clear gap in the singular values for 
{\tt WNNopt}. The difference in detected order despite similar 
VAF indicates that order detection is not so
critical for this example. 


\subsubsection{The Hair dryer data set (\# 3 in
  Table~\ref{tab:DAISYdescr})}

\begin{figure}[h]
\includegraphics[width=\linewidth]{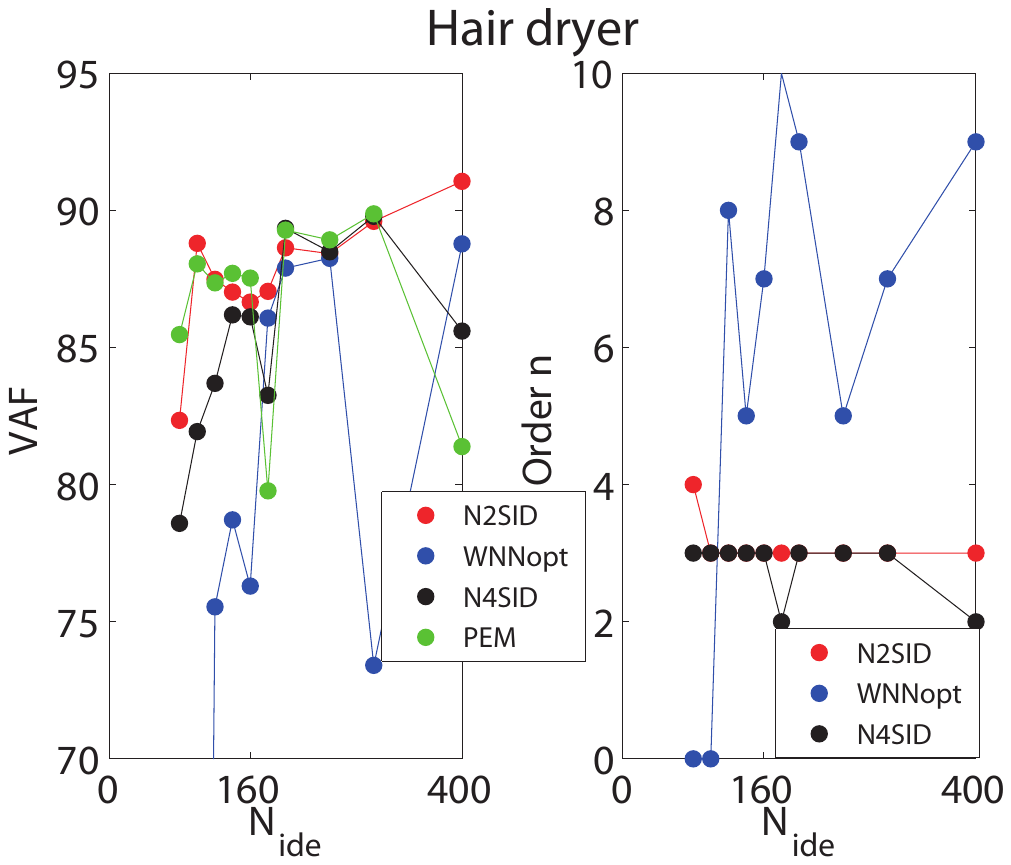}
\caption{VAF Daisy \# 3 - Hair dryer.}
\label{fig:vaf_daisy3}
\end{figure}

The goodness of fit parameter VAF and the
estimated model order $\hat{n}$ are plotted in
Figure~\ref{fig:vaf_daisy3} in the left and right graphs, respectively. Here it
is again clear that {\tt N2SID} outperforms all other SID methods and
provides more stable behavior when increasing $N_{\rm ide}$ compared
to the fluctuating behavior of the other methods, both with respect to VAF and
estimated model order. {\tt WNNop} fails to address the case of
very small data length sets, i.e. when
$N_{\rm ide} = 80$ and $100$. The combination of {\tt
  N4SID} and {\tt PEM} enables in a number of cases to provide models
with a similar VAF compared to {\tt N2SID} and in a small number of
cases  to slightly improve the results over {\tt N2SID}. However,
this is not consistent, since for $N_{\rm ide} = 400$ the VAF
is worsened compared to the intialization with {\tt N4SID}.

\begin{figure}
\includegraphics[width=\linewidth]{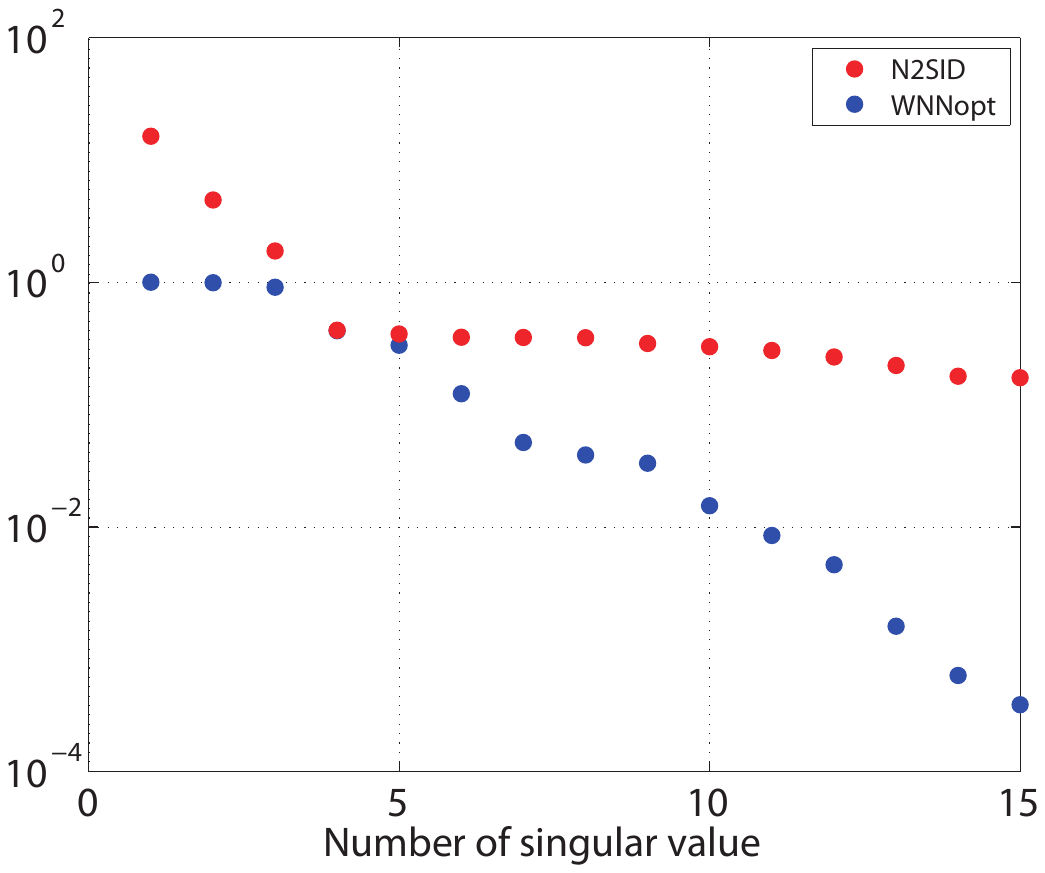}
\caption{Singular values Daisy \# 3 - Hair dryer.}
\label{fig:sv_daisy3}
\end{figure}

Figure~\ref{fig:sv_daisy3} diplays the singular values for the
last data set where $N_{\rm ide} = 400$ in Figure~\ref{fig:sv_daisy3}. It
confirms the improved potential in low rank approximation by
{\tt N2SID} over {\tt WNNopt}. The latter method diminishes the gap,
leading in general to a larger model order
estimation. This larger model order does however for this example 
not lead to a better output prediction. 

\subsubsection{The Steam Heat Exchanger data set (\# 4 in
  Table~\ref{tab:DAISYdescr})}

\begin{figure}[h]
\includegraphics[width=\linewidth]{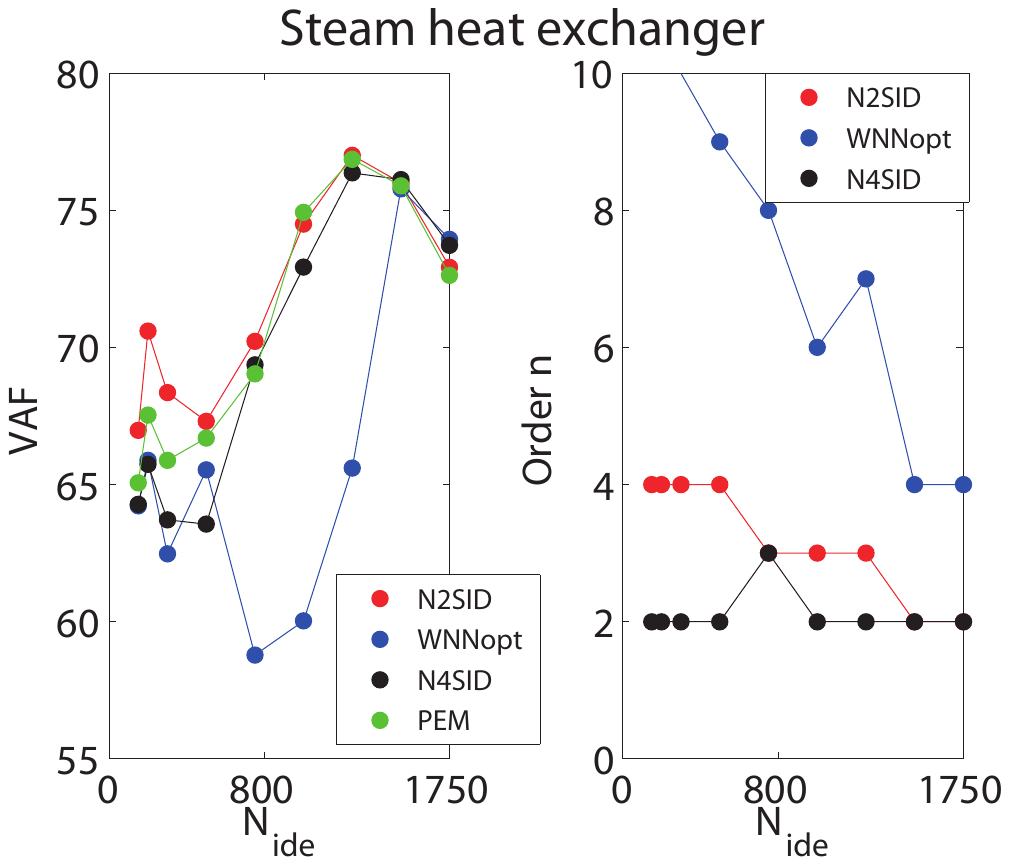}
\caption{VAF Daisy \# 4 - Steam Heat Exchanger.}
\label{fig:vaf_daisy4}
\end{figure}

The goodness of fit parameter VAF and the
estimated model order $\hat{n}$ are plotted in
Figure~\ref{fig:vaf_daisy4}.  

{\tt N2SID} again for small data sets with $N_{\rm ide}$ ranging
between $150$ and $750$ yields the best output predictions of all
methods.  
Comparing the
VAF value in Figure~\ref{fig:vaf_daisy4} with those for the
previous Daisy data sets reveals that the  values are smaller. This
reflects problems with the data set due to lower signal to noise
ratio, system nonlinearity, etc. Because of this
we started the analysis with the smallest value of $N_{\rm ide}$ equal
to $150$, since for smaller values poor results were obtained for all
methods. 

The other methods show a similar behavior as for the previously
analysed data sets: in most cases but not all {\tt PEM} improves over
{\tt N4SID}, {\tt WNNopt} displays inferior behavior for 
$N_{\rm ide} \leq 1250$, and both {\tt N2SID} and {\tt N4SID (PEM)}
determine a smaller order then {\tt WNNopt}.

\begin{figure}
\includegraphics[width=\linewidth]{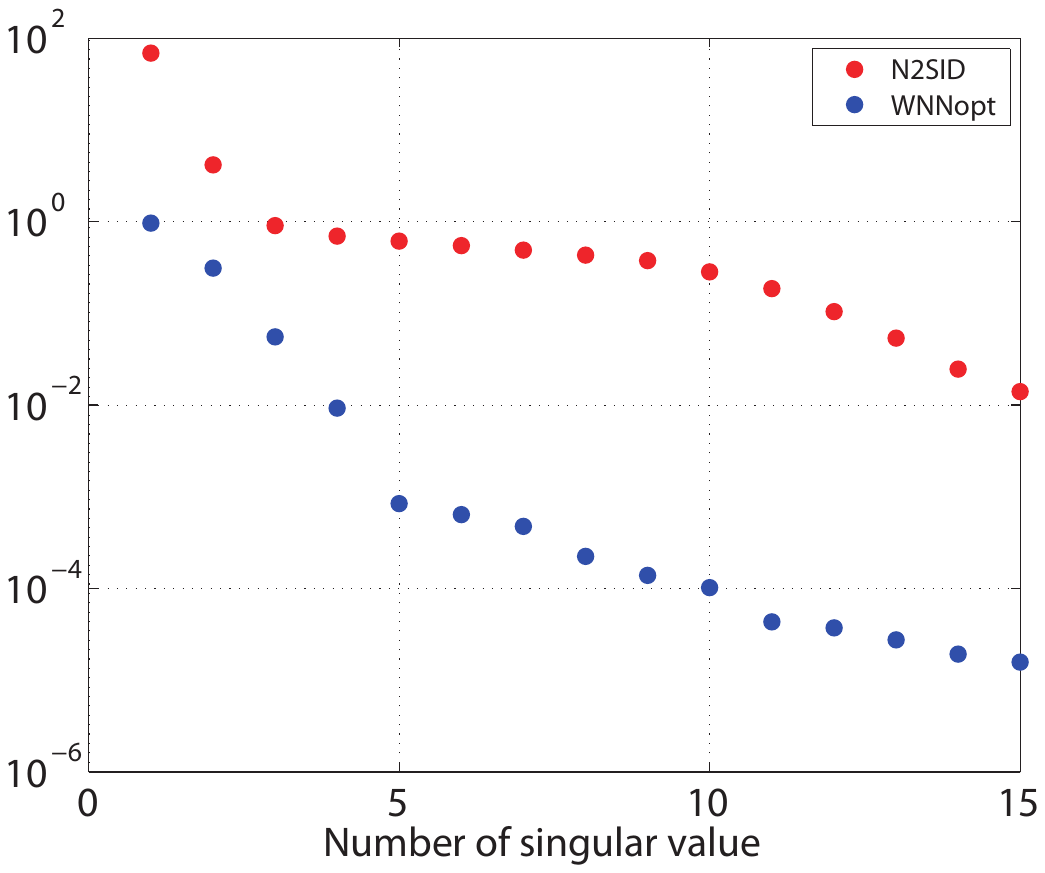}
\caption{Singular values Daisy \# 4 - Steam Heat Exchanger.}
\label{fig:sv_daisy4}
\end{figure}

Finally, the plot of the singular values for the last data set in
Figure~\ref{fig:sv_daisy4} displays a similar behavior. 
Both singular value plots clearly support the automatic order
selection made. However {\tt N2SID} has a better trade-off
between model complexity and model accuracy as expressed by the VAF. 


\subsubsection{Heat flow density data set (\# 5 in
  Table~\ref{tab:DAISYdescr})}

\begin{figure}[h]
\includegraphics[width=\linewidth]{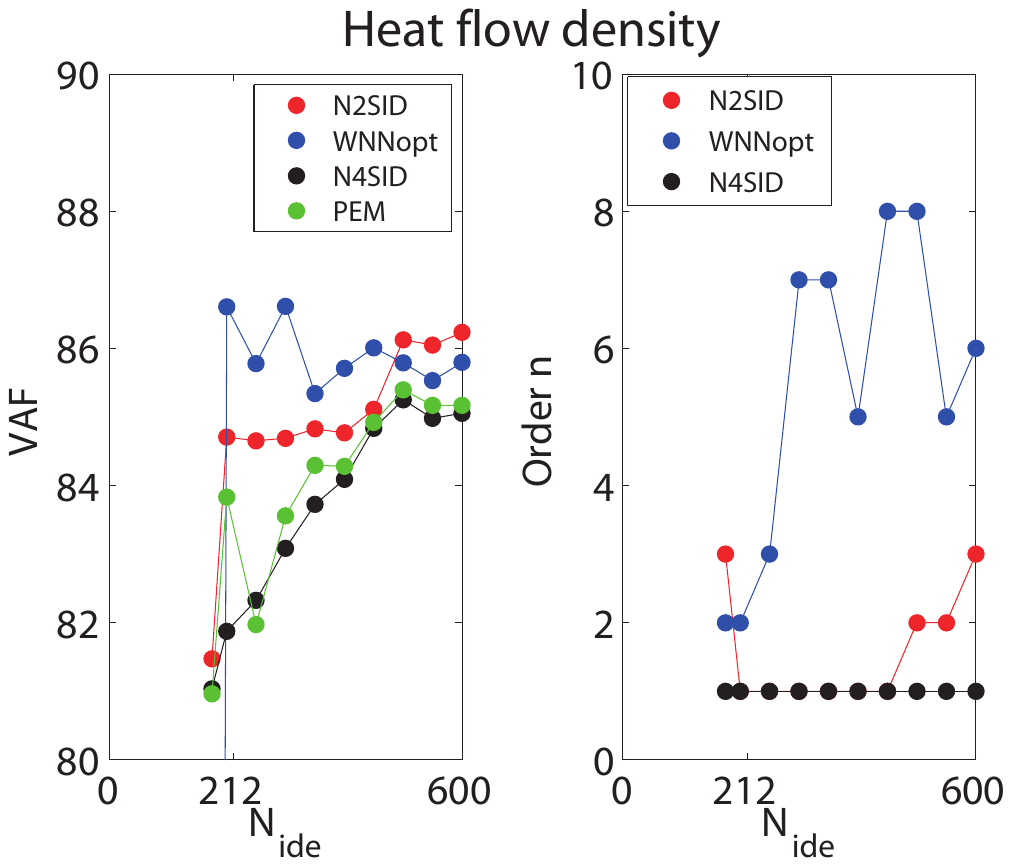}
\caption{VAF Daisy \# 5 -Heat flow density.}
\label{fig:vaf_daisy7}
\end{figure}

The goodness of fit parameter VAF and the
estimated model order $\hat{n}$ are plotted in
Figure~\ref{fig:vaf_daisy7}.  

For this data set {\tt WNNopt} provides for $200 \leq N_{\rm ide} \leq
450$ the best results but in general detects a larger model order.
For the smallest length data set {\tt WNNopt} produced inferior
VAF. {\tt N2SID} provides a better VAF prediction compared to {\tt
  N4SID} and {\tt PEM} and this for a smaller model order $\hat{n}$ as
compared to {\tt WNNopt}. 

\begin{figure}
\includegraphics[width=\linewidth]{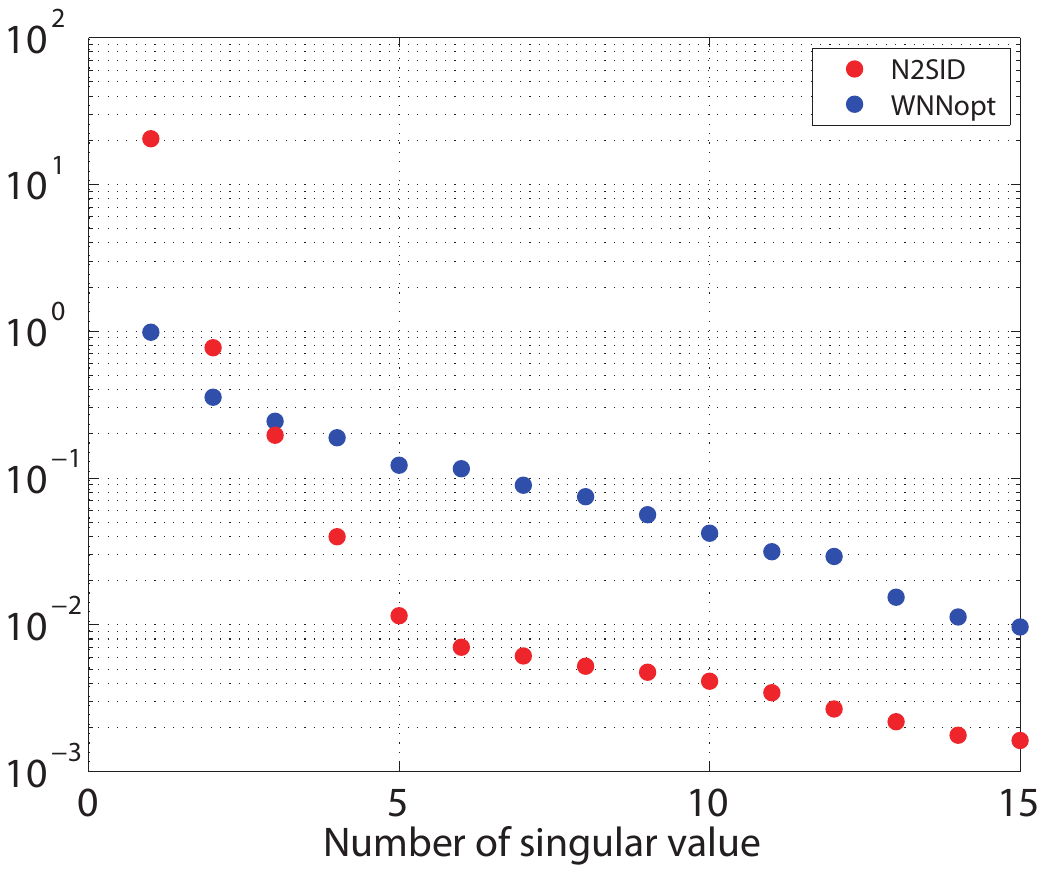}
\caption{Singular values Daisy \# 5 - Heat flow density.}
\label{fig:sv_daisy7}
\end{figure}

From the singular values in  Figure~\ref{fig:sv_daisy7} for $N_{\rm
  ide} = 600$ an order selection of $1$ up to $4$ is clearly justified
by {\tt N2SID}. The order selection made by {\tt WNNopt} is much less
clear. 
\subsubsection{General Observation from the analysed Daisy data sets.}  

The automized analysis of the $5$ Daisy data sets clearly demonstrates
the merit of the new SID method {\tt N2SID} over the other representative
identification methods considered. Especially when
considering data sets of {\em small length} it is able to make a good
and sometimes excellent trade-off between model complexity and
model accuracy as expressed by the goodness of fit. The improvement over the
other analysed nuclear norm subspace identification method {\tt
  WNNopt} in
order detection both in revealing a clear gap as well as in detecting
models of low complexity is evident. 

\section{Concluding Remarks}

Subspace identification of multivariable state space innovation models 
is revisited in this paper in the scope of
nuclear norm optimization methods and using the observer form. A new subspace identification
method is presented, referred to as N2SID. N2SID is the first subspace
identification method that addresses the identification of innovation
state space models {\em without} the use of instrumental
variables (IVs). The avoidance of using IVs leads to a number of
improvements. First as shown in the experimental study in
\cite{Report}, it leads to improved results in identifying innovation models
when compared to existing SID methods, like N4SID and 
the recent Nuclear Norm based SID methods
presented in \cite{liu+han+van13} and with the Prediction Error Method
(PEM) \cite{Ljungtb:07}. This improvement especially holds for small
length data batches, i.e. when the number of samples is only a small
multiple of the order of the underlying system. Second, as illustrated
by Theorem 1, the methodology presented enables to provide insight on
the necessary conditions of persistency of excitation of the input on
the existance of a unique solution. Finally, the new N2SID methodology
will enable to address other interesting identification problems in a
subspace identification framework, such as the identification of
distributed systems as shown in \cite{IFAC1,IFAC2,IFAC3}.  \\[2pt]
{\footnotesize \em Acknowledgement: The authors kindly acknowledge Mr. Baptiste
  Sinquin from Ecole Centrale Lyon for his help in a preliminary
  matlab comparison study with N2SID 
 during his internship 
  at the Delft Center for Systems and Control under the supervision of
  Prof. M. Verhaegen. Also the discussions with Dr. Chengpu Yu of the
  Delft Center for Systems and Control on the
  topic of Theorem 1 are very much appreciated.} \\[3pt]
{\footnotesize \em The authors express their full appreciation for the
  constructive comments raised by the anonymous reviewers.}
\bibliographystyle{plain}        
\bibliography{refN2SID,cvx-refs2,database2,myref,subspace}

\section*{Appendix: ADMM Algorithm}
We here state the ADMM algorithm for a generic
nuclear norm optimization problem with a quadratic regularization term:
\BEQ \label{e-na}
\begin{array}{ll}
  \mbox{minimize} & \displaystyle 
 \|\mathcal A(x) + A_0 \|_* + \frac{1}{2} (x-a)^T H (x -a).
\end{array}
\EEQ
The presentation is an allmost exact citation from \cite{liu+han+van13}.
The variable is a vector $x\in\reals^n$.  
The first term in the objective is the nuclear norm of a
$p\times q$ matrix $\mathcal A(x) + A_0$ where
$\mathcal A:\reals^n \rightarrow \reals^{p\times q}$ is a linear mapping.
The parameters in the second, quadratic, term 
in the objective of~(\ref{e-na}) are a vector $a\in \reals^n$ and 
a positive semidefinite matrix  $H \in\symm^n$.

To derive the ADMM iteration we first write~(\ref{e-na}) as
\[
  \begin{array}{ll}
  \mbox{minimize}   & \|X\|_* + (1/2)(x-a)^T H (x- a) \\
  \mbox{subject to} & \mathcal A(x) + A_0 = X 
 \end{array}
\]
with two variables $x\in\reals^n$ and $X\in\reals^{p\times q}$.
The \emph{augmented Lagrangian} for this problem is
\BEA
 L_\rho(x, X, Z) & =& \|X\|_* + \frac{1}{2}(x-a)^T H (x-a) \\
&+& \Tr(Z^T(\mathcal A(x) + A_0- X))\\
&+& \frac{\rho}{2} \|\mathcal A(x) + A_0 - X\|^2_F,
\EEA
where $\rho$ is a positive penalty parameter.  
Each iteration of the ADMM consists of a minimization of
$L_\rho$ over $x$, a minimization of $L_\rho$ over $X$, and a simple
update of the  dual variable $Z$.  This is summarized 
in Table~\ref{table:ADMM}.
\begin{table}
\caption{ADMM algorithm \label{table:ADMM}}
\begin{quote}
\begin{tabbing}
1.\  \= Initialize $x$, $X$, $Z$, $\rho$.  For example, \\set
  $x=0$, $X=A_0$, $Z=0$, $\rho=1$. 
  \\*[\smallskipamount]
2.\> Update $x: = \argmin_{\hat x} L_\rho(\hat x, X, Z)$. 
  See~(\ref{e-ax}). 
  \\*[\smallskipamount]
3.\> Update $X:= \argmin_{\hat X} L_\rho(x, \hat X, Z)$. 
  See~(\ref{e-aY}). 
  \\*[\smallskipamount]
4.\> Update $Z := Z + \rho (\mathcal A(x) + A_0 - X)$.
  \\*[\smallskipamount]
5.\> Terminate if
  $\|r_\mathrm{p}\|_F \leq \epsilon_\mathrm{p}$
and $\|r_\mathrm{d}\|_2 \leq \epsilon_\mathrm{d}$  
 (see~(\ref{e-rp})--(\ref{e-ed})).\\  
 Otherwise, go to step 2. 
\end{tabbing} 
\end{quote}
\end{table}

The update in step~2 requires the solution of a linear equation,
since $L_\rho(\hat x, X, Z)$ is  quadratic in $\hat x$.  Setting
the gradient of $L_\rho(\hat x,X,Z)$ with respect to $\hat x$ equal
to zero gives the equation
\BEQ \label{e-ax}
 (M + \rho H)\hat x 
 = \mathcal A_\mathrm{adj}(\rho X + \rho A_0 - Z) + Ha
\EEQ
where $\Aa$ is the adjoint of the mapping $\mathcal A$
and $M$ is the positive semidefinite matrix defined by the identity 
\BEQ \label{e-M}
 Mz = \Aa(\mathcal A(z)) \quad \forall z.
\EEQ

The minimizer $X$ in step~4 is obtained by soft-thresholding the singular 
values of the matrix $\mathcal A(x) + A_0 + Z/\rho$:
\BEQ \label{e-aY}
\argmin_{\hat X} L_\rho(\hat X,x,Z) = \sum_{i=1}^{\min\{p,q\}}
 \max\{0, \sigma_i - \frac{1}{\rho}\} \; u_i v_i^T
\EEQ
where $u_i$, $v_i$, $\sigma_i$ are given by a singular value decomposition
\[
 \mathcal A(x) + A_0 + \frac{1}{\rho}Z = \sum_{i=1}^{\min\{p,q\}}
 \sigma_i u_i v_i^T.
\]
The residuals and tolerances in the stopping criterion 
in step~5 are defined as follows \cite{BPCPE:11}:
\BEA
 r_\mathrm{p} & = & \mathcal A(\mathbf x) + A_0 - X \label{e-rp}\\
 r_\mathrm{d} & = & \rho \Aa (X_\mathrm{prev} - X) \label{e-rd} \\
 \epsilon_\mathrm{p} & = & \sqrt{pq} \, \epsilon_\mathrm{abs} +
\epsilon_\mathrm{rel} \max \{\|\mathcal A(x)\|_F, \|X\|_F, 
\|\mathcal A_0\|_F \} \label{e-er} \\
 \epsilon_\mathrm{d} & = & \sqrt{n} \epsilon_\mathrm{abs} +
\epsilon_\mathrm{rel} \, \| \Aa (Z) \|_2, \label{e-ed}
\EEA
Typical values for the relative and absolute tolerances are 
$\epsilon_\mathrm{rel} = 10^{-3}$
and $\epsilon_\mathrm{abs} = 10^{-6}$.  
The matrix $X_\mathrm{prev}$ in~(\ref{e-rd}) is the value of $X$ in 
the previous iteration.

Instead of a using a fixed penalty parameter $\rho$, 
one can vary $\rho$ to improve the speed of convergence.
An example of such a scheme is to adapt $\rho$ at the
end of each ADMM iteration as follows~\cite{HYW:00} 
\[
 \rho := \left\{ 
  \begin{array}{ll}
   \tau \rho \quad & \|r_\mathrm{p}\|_F > \mu \|r_\mathrm{d}\|_2 \\
   \rho/\tau \quad & \|r_\mathrm{d}\|_2 > \mu \|r_\mathrm{p}\|_F \\
   \rho \quad & \mbox{otherwise.}
  \end{array} \right.
\]
This scheme depends on parameters $\mu >1$, $\tau>1$ 
(for example, $\mu = 10$ and $\tau=2$).
Note that varying $\rho$ has an important consequence on the 
algorithm in Table~\ref{table:ADMM}.   
If $\rho$ is fixed, the coefficient matrix $H + \rho M$ in the 
equation~(\ref{e-ax}) that is solved in step~2 of each iteration 
is constant throughout the algorithm.  
Therefore only one costly factorization of $H+\rho M$ is required.
If we change $\rho$ after step~6, a new factorization of
$H+\rho M$ is needed before returning to step~3.
I is explain in \cite{liu+han+van13} how the extra cost of 
repeated factorizations can be avoided.

\end{document}